\documentclass[conference]{IEEEtran}

\usepackage[english]{babel}
\usepackage[utf8]{inputenc}
\usepackage{cite}
\usepackage[pdftex]{graphicx}
\usepackage{amsmath}
\usepackage{amssymb}
\usepackage{algorithmic}
\usepackage{array}
\usepackage{url}
\usepackage{tcolorbox}
\usepackage{multirow}
\usepackage{xcolor}
\usepackage{booktabs}
\usepackage[caption=false]{subfig}
\usepackage[normalem]{ulem}
\usepackage{balance}

\DeclareGraphicsExtensions{.pdf,.jpeg,.png}

\newcommand{\step}[1]{\vspace{1mm} \noindent {\em #1:}}

\usepackage{changepage}

\usepackage{framed}
\makeatletter
  {\par\unskip\endMakeFramed}
\makeatother

\definecolor{formalshade}{rgb}{0.93,0.93,0.93}

\definecolor{darkblue}{rgb}{0.2, 0.2, 0.2}

\newenvironment{formal}{%
  \def\FrameCommand{%
    \hspace{1pt}%
    {\color{darkblue}\vrule width 2pt}%
    {\color{formalshade}\vrule width 4pt}%
    \colorbox{formalshade}%
  }%
  \MakeFramed{\advance\hsize-\width\FrameRestore}%
  \noindent\hspace{-1pt}
  \begin{adjustwidth}{}{7pt}%
  \vspace{2pt}\vspace{2pt}%
}
{%
  \vspace{3pt}\end{adjustwidth}\endMakeFramed%
}

\newcounter{resultcounter}

\newcounter{patterncounter}

\makeatletter
\def\@xfootnote[#1]{%
  \protected@xdef\@thefnmark{#1}%
  \@footnotemark\@footnotetext}
\makeatother

\begin{document}

\title{
  Identifying Experts in Software Libraries and Frameworks among GitHub Users
}
  
\author{\IEEEauthorblockN{João Eduardo Montandon}
\IEEEauthorblockA{
Technical College (COLTEC)\\
Federal University of Minas Gerais\\
Belo Horizonte, Brazil\\
{\em joao.montandon@dcc.ufmg.br}
}
\and
\IEEEauthorblockN{Luciana Lourdes Silva}
\IEEEauthorblockA{
Department of Computer Science\\
Federal Institute of Minas Gerais \\
Ouro Branco, Brazil\\
{\em luciana.lourdes.silva@ifmg.edu.br}
}
\and
\IEEEauthorblockN{Marco Tulio Valente}
\IEEEauthorblockA{
Department of Computer Science\\
Federal University of Minas Gerais\\
Belo Horizonte, Brazil\\
{\em mtov@dcc.ufmg.br}
}
}
  
  \maketitle
  
  \newcommand{\react}{{\sc facebook/react}}
  \newcommand{\mongodb}{{\sc mongodb/node-mongodb}}
  \newcommand{\socketio}{{\sc socketio/socket.io}}

  \newcommand{\sreact}{{\sc react}}
  \newcommand{\smongodb}{{\sc node-mongodb}}
  \newcommand{\ssocketio}{{\sc socket.io}}

 \begin{abstract}
Software development increasingly depends on libraries and frameworks to increase productivity and reduce time-to-market. Despite this fact, we still lack techniques to assess developers expertise in widely popular libraries and frameworks. In this paper, we evaluate the performance of unsupervised (based on clustering) and supervised machine learning classifiers (Random Forest and SVM) to identify experts in three popular JavaScript libraries: facebook/react,  mongodb/node-mongodb, and socketio/socket.io. First, we collect 13 features about developers activity on GitHub projects, including commits on source code files that depend on these libraries. We also build a ground truth including the expertise of 575 developers on the studied libraries, as self-reported by them in a survey. Based on our findings, we document the challenges of using machine learning classifiers to predict expertise in software libraries, using features extracted from GitHub. Then, we propose a method to identify library experts based on clustering feature data from GitHub; by triangulating the results of this method with information available on Linkedin profiles, we show that it is able to recommend dozens of GitHub users with evidences of being experts in the studied JavaScript libraries. We also provide a public dataset with the expertise of 575 developers on the studied libraries.

 \end{abstract}
  
  \section{Introduction}
  \label{sec:Introduction}
  
  

Modern software development heavily depends on libraries and frameworks to increase productivity and reduce time-to-market~\cite{Ruiz2014,Sawant2017}. In this context,
identifying experts in popular libraries and frameworks---for example, among the members of global open-source software development platforms, like GitHub---has a practical value. 
For example, open source project managers can use this information to search for potential new contributors to their systems. Private companies can also benefit from this information before hiring developers to their projects. In fact, we manually inspected 1,839 job offers, available on July 2nd, 2018 at Stack Overflow Jobs.\footnote{https://stackoverflow.com/jobs} We found that 789 jobs (42\%) have at least one tag referring to  frameworks and libraries, including {\sc ReactJS} (372 jobs), {\sc AngularJS} (215 jobs), and {\sc Ruby on Rails} (135 jobs). This result suggests that companies, when hiring, often target developers with expertise in specific programming technologies. Furthermore, this information can help to recommend experts to answer questions in Q\&A forums~\cite{Treude2011} or to assist project managers to set up balanced development teams~\cite{Siau2010}.

Previous work on software expertise focused on identifying experts for internal parts of a software project, but not on external components, such as libraries and frameworks. For example, Expertise Browser~\cite{Mockus2002b} visually maps parts of a software product (e.g., code or documentation) to the respective experts, using number of changes (commits) as the basic measure of expertise.
Fritz et al.~\cite{Fritz2007,Fritz2010,Fritz2014} propose the degree-of-knowledge (DOK) metric to identify experts in specific source-code files, which combines both commits and interactions with the code, by means of an IDE.
Schuler and Zimmerman~\cite{Schuler2008} advocate that expertise can also be gained by using the component of interest (e.g., by calling its methods). Silva-Junior et al.~\cite{DaSilva2015} propose a fine-grained approach to identify expertise in specific source-code elements---methods, classes, or packages. However, these works aim to identify experts that can fix a bug, review or evolve  internal parts of an specific software product. 

In this paper, we extend existing expertise identification approaches to the context of third-party software components. Our key hypothesis is that when maintaining a piece of code, developers also gain expertise on the frameworks and libraries used by its implementation. We focus on three popular libraries: {\sc facebook/react} (for building enriched Web interfaces), {\sc mongodb/node-mongodb} (for accessing MongoDB databases), and {\sc socketio/socket.io} (for real-time communication). Then, we evaluate the use of unsupervised (based on clustering) and supervised machine learning classifiers to identify experts in these libraries.
Both techniques are applied using features about candidate experts in each library, extracted for selected GitHub users. These features include, for example, number of commits on files that import each library and number of client  projects a candidate expert has contributed to. We also survey a sample of GitHub users to create a ground truth of developers expertise in the studied libraries. In this survey, the participants declared their expertise (in a scale from 1 to 5) in the libraries. This ground truth provides the expertise of 575 GitHub developers in the studied libraries, including 418 \react\ developers, 68 \mongodb\ developers, and 89 \socketio\ developers. To validate our hypothesis, we first train and evaluate two  machine learning classifiers, based on Random Forest~\cite{Breiman2001} and SVM~\cite{Weston1998}. Finally,
we investigate the use of clustering algorithms  to 
identify library experts.
  
\step{Research Questions} We ask two research questions:\\[-0.3cm]
  
  \noindent {\em (RQ.1) How accurate are machine learning classifiers in identifying library experts?} For three expertise classes---novices, intermediate, and experts---the maximal F-measure is 0.56 ({\sc mongodb/node-mongodb}).
We argue that this poor performance is inherent of using GitHub as a full proxy for expertise. For example, there are   experts that rarely contribute to public GitHub projects; their expertise comes from working on private projects or projects that are not GitHub-based.
low feature values (e.g., commits in library clients), making it  challenging to predict the expertise of such developers, by considering their activity on GitHub.\\[-0.3cm]
 
\noindent{\em (RQ.2) Which features best distinguish experts in the studied libraries?} In this second {\em RQ}, we first rely on clustering to identify experts that share similar feature values. In \react, we found a cluster where 74\% of the developers are experts in the framework; in \mongodb\ and \socketio\, we found clusters with 65\% and 75\% of experts, respectively. More importantly, we show that the experts in such clusters tend to be active and frequent contributors to library clients on GitHub. Therefore, this finding suggests that GitHub data can be a partial proxy for expertise in libraries and frameworks. By partial proxy, we mean that developers with high feature values (commits, code churn, etc) tend to be experts in the studied libraries; by contrast, the proxy fails in the case of developers with low feature values, who can be both experts and novices, as concluded in {\em RQ.1}.

\step{Contributions} Our contributions are threefold: (1) based on the findings and lessons learned with {\em RQ.1}, we document the challenges of using machine learning classifiers to predict expertise in software libraries, using features extracted from GitHub; (2) inspired by the findings of {\em RQ.2}, we propose an unsupervised method to identify library experts based on clustering feature data from GitHub; by triangulating the results of this method with expertise information available on Linkedin, we show that it is able to recommend dozens of GitHub users with robust evidences of being experts in \react, a popular JavaScript library; (3) we provide a public ground truth with the expertise of 575 developers on three relevant JavaScript libraries; to our knowledge, this is the largest dataset with expertise data on specific software technologies.
 
\step{Structure} Section \ref{sec:study_design} documents the process we followed to collect the data used to answer {\em RQ.1} and {\em RQ.2}. Section \ref{sec:methods} des\-cribes the techniques used in this work, as well as their setup. Section \ref{sec:results} provides answers to the proposed {\em RQs}. Section \ref{sec:discussion} summarizes our findings, lessons learned, and limitations. It also proposes a practical method for identifying library experts and validates its results with Linkedin data. Section \ref{sec:threats_validity} reports threats to validity and Section \ref{sec:related_work} describes related work. Finally, Section \ref{sec:conclusion} concludes the paper.

\section{Data Collection}
  \label{sec:study_design}
  
  \subsection{Definitions}
  
  Before presenting the data collection process, we define key terms used in this process and also in the rest of this paper:
  
  \begin{itemize}
  
  \item {\em Target Library}: The JavaScript libraries used in this paper; our goal is to  identify experts in these libraries based on their activity on GitHub.
  
  \item {\em Client Project (or File)}: A project (or source code file) that depends on a target library.
  
  \item {\em Candidate Expert:} A contributor of a client project whose expertise on a target library is assessed in this paper.
  
  \item {\em Feature:} An attribute of a candidate expert that may act as a predictor of its expertise on a target library.
  
  \item {\em Ground Truth:} A dataset with the expertise of candidate experts in a target library, as self-reported by them.
  \end{itemize}

  \subsection{Target Libraries}
  
We evaluate JavaScript libraries due to the importance and popularity of this language in modern software development. 
We focus on the developers of three JavaScript libraries\footnote{In our study, the terms libraries and frameworks are used interchangeably.}: {\sc facebook/react}\footnote{\url{https://github.com/facebook/react}} (a system for building enriched Web interfaces), {\sc mongodb/node-mongodb}\footnote{\url{https://github.com/mongodb/node-mongodb-native}} (the official Node.js driver for MongoDB database server), and {\sc socketio/socket.io}\footnote{\url{https://github.com/socketio/socket.io}} (a library for real-time communication). 
  We selected {\sc facebook/react} because it is a very popular  front-end development library; after making this first selection, we searched for libraries handling important concerns in back-end development and selected {\sc mongodb/node-mongodb}, a persistence library; and {\sc socketio/socket.io}, since communication is important both in front-end and back-end programming. 
  Table~\ref{tab:libraries} shows information about these systems, including number of stars, contributors, commits, and files (on April, 2018). As we can see, they are popular projects (at least 6,696 stars) and actively maintained (at least 149 contributors and 1,698 commits). For brevity, we call them \sreact, \smongodb, and \ssocketio\ in the rest of this paper.

  \begin{table}[!t]
  \centering
  \caption{Target Libraries}
  \label{tab:libraries}
  \begin{tabular}{lrrrr}
  \toprule
  {\bf Target Library} & {\bf Stars} & {\bf Contrib} & {\bf Commits} & {\bf Files} \\ 
  \midrule
  \react\ & 91,739 & 1,171 & 9,731 & 797 \\
  \mongodb\ & {6,696} & {260} & {4,565} & {617} \\
  \socketio\ & 40,199 & 149 & 1,698 & 83 \\
  \bottomrule
  \end{tabular}
  \end{table} 

 \setcounter{table}{2}
   \begin{table*}[!t]
  \centering
  \caption{Features collected for each candidate expert in each target library}
  \label{tab:features}
  \begin{tabular}{lll}
  \toprule
  \textbf{Dimension}        & \textbf{Feature}          & \textbf{Description}                                                                                   \\ \midrule
  \multirow{6}{*}{Volume}  & commits                     & Number of commits in client projects                                                                     \\
                             & commitsClientFiles        & Number of commits changing at least one client file                                                      \\
                             & commitsImportLibrary        & Number of commits adding library import statements                                                       \\
                             & codeChurn                  & Code churn considering all commits in client projects                                                    \\
                             & codeChurnClientFiles   & Code churn considering only changes in client files                                                      \\
                             & imports           & Number of added library import statements                                                                \\ \midrule
  \multirow{5}{*}{Frequency} & daysSinceFirstImport         & Number of days since the first commit where a library import statement was added                         \\
                             & daysSinceLastImport         & Number of days since the last commit where a library import statement was added                          \\
                             & daysBetweenImports          & Number of days between the first/last commits where a library import statement was added                 \\
                             & avgDaysCommitsClientFiles & Average interval (in days) of the commits changing client files                                          \\
                             & avgDaysCommitsImportLibrary & Average interval (in days) of the commits adding library import statements                               \\ \midrule
  \multirow{2}{*}{Breadth} & projects                    & Number of client projects the developer contributed at least once                                        \\
                             & projectsImport            & Number of client projects where the developer added a library import statement                           \\ \bottomrule
  
\end{tabular}
\end{table*}

  \subsection{Candidate Experts} 
  \label{sec:experts}
  
  For each target library $\mathcal{L}$, where $\mathcal{L}$ is \sreact, \smongodb, or \ssocketio, we selected a list of candidate experts, as described next.
  First,  we relied on the top-10K most popular JavaScript projects on GitHub, according to their number of stars. We checked out these projects and  searched for dependencies to $\mathcal{L}$ in {\em package.json} and {\em bower.json} files, which are configuration files used by two popular JavaScript package managers.
  A candidate expert in $\mathcal{L}$ is a developer who performed at least one change in a source code file (from a client project) that depends on $\mathcal{L}$. In other words, we assume that if a developer changed a file that imports $\mathcal{L}$ he has chances to be an expert in this library. Next, we removed aliases from this initial list of candidate experts, i.e.,~the same developer, but with distinct e-mails on the considered commits. For this purpose, we used a feature of GitHub API that maps a commit author to its GitHub account. Using this feature, we mapped each developer in the list of candidate experts to his/her GitHub's account. Candidate experts $e$ and $e'$ are the same when they share the same GitHub account. Table \ref{tab:candidate_experts} shows for each target library the number of client projects, and the final number of candidate experts after handling aliases. As we can observe, \sreact\ has the highest number of both client projects (1,136) and candidate experts (8,742). Therefore, our dataset includes a popular target library, with thousands of client projects and candidate experts; but it also includes less popular libraries, with just a few hundred candidate experts.

 
 \setcounter{table}{1} 
 \begin{table}[!h]
  \centering
  \caption{Client Projects and Candidate Experts}
  \label{tab:candidate_experts}
  \begin{tabular}{lrrr}
  \toprule
  {\bf Library} & {\bf Clients} & {\bf Experts} \\ 
  \midrule
  \react\ & 1,136 & 8,742\\
  \mongodb\ & 223 & 454 \\
  \socketio\ & 345 & 608 \\
  \bottomrule
  \end{tabular}
  \end{table}
   \setcounter{table}{3} 
 
\subsection{Features}
  
We collected 13 features for each candidate expert selected in the previous step. As documented in Table \ref{tab:features}, these features cover three dimensions of {\em changes} performed on client files.\footnote{These dimensions and their features were derived and extended from the literature on developers expertise in open source communities. Volume of changes (particularly, number of commits) is commonly used in related works~\cite{Mockus2002b,Fritz2007,Fritz2010,Fritz2014}. Frequency and breadth of changes have also been considered as proxies to developers expertise~\cite{DabbishSTH12, ieeesw-2019,icpc2016,Singer2013, Marlow2013, DaSilva2015}. As an additional criterion, we only use features that can be directly computed from GitHub public API.} 
  
\begin{itemize}
  \item {\em Volume of changes}, which includes six features about the quantity of changes performed by candidate experts in client projects, such as  number of commits and code churn (e.g.,~lines added or deleted). We conjecture that by heavily maintaining a file developers gain expertise on libraries used by its implementation.
  
  \item {\em Frequency of changes}, including five features expressing the frequency and time of the changes performed by candidate experts, e.g., number of days since first and last library import. The rationale is that expertise also depends on temporal properties of the changes.
  
  \item {\em Breadth of changes}, which includes two features about the number of client projects the candidate experts worked on. The rationale is that expertise might increase when candidate experts work in different client projects.
\end{itemize}
  
\begin{figure*}[!t]
\centering
\subfloat[ref1][\react]{\includegraphics[width=0.3\textwidth]{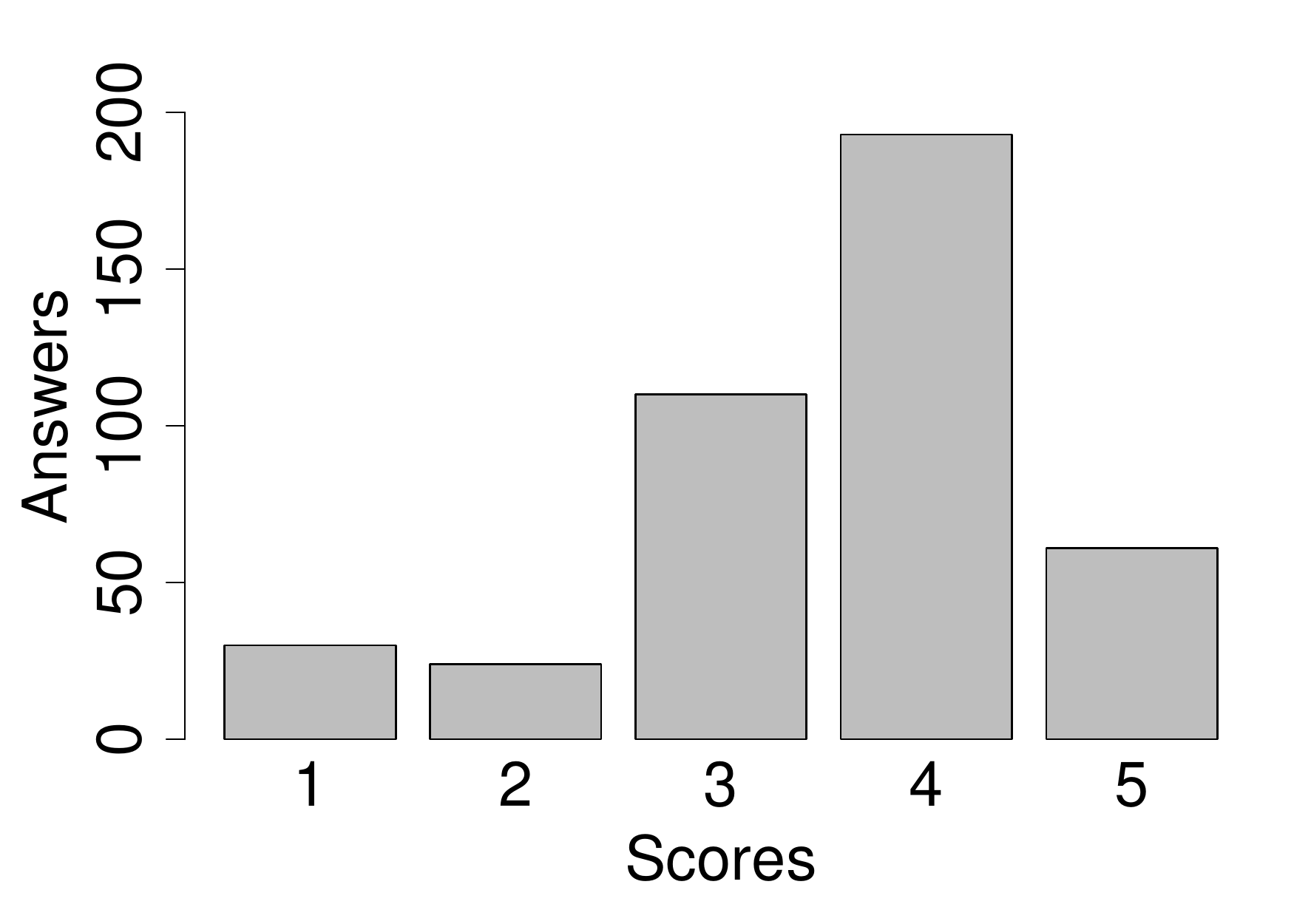}}
\quad
\subfloat[ref2][{\sc mongodb/mongo}]{\includegraphics[width=0.3\textwidth]{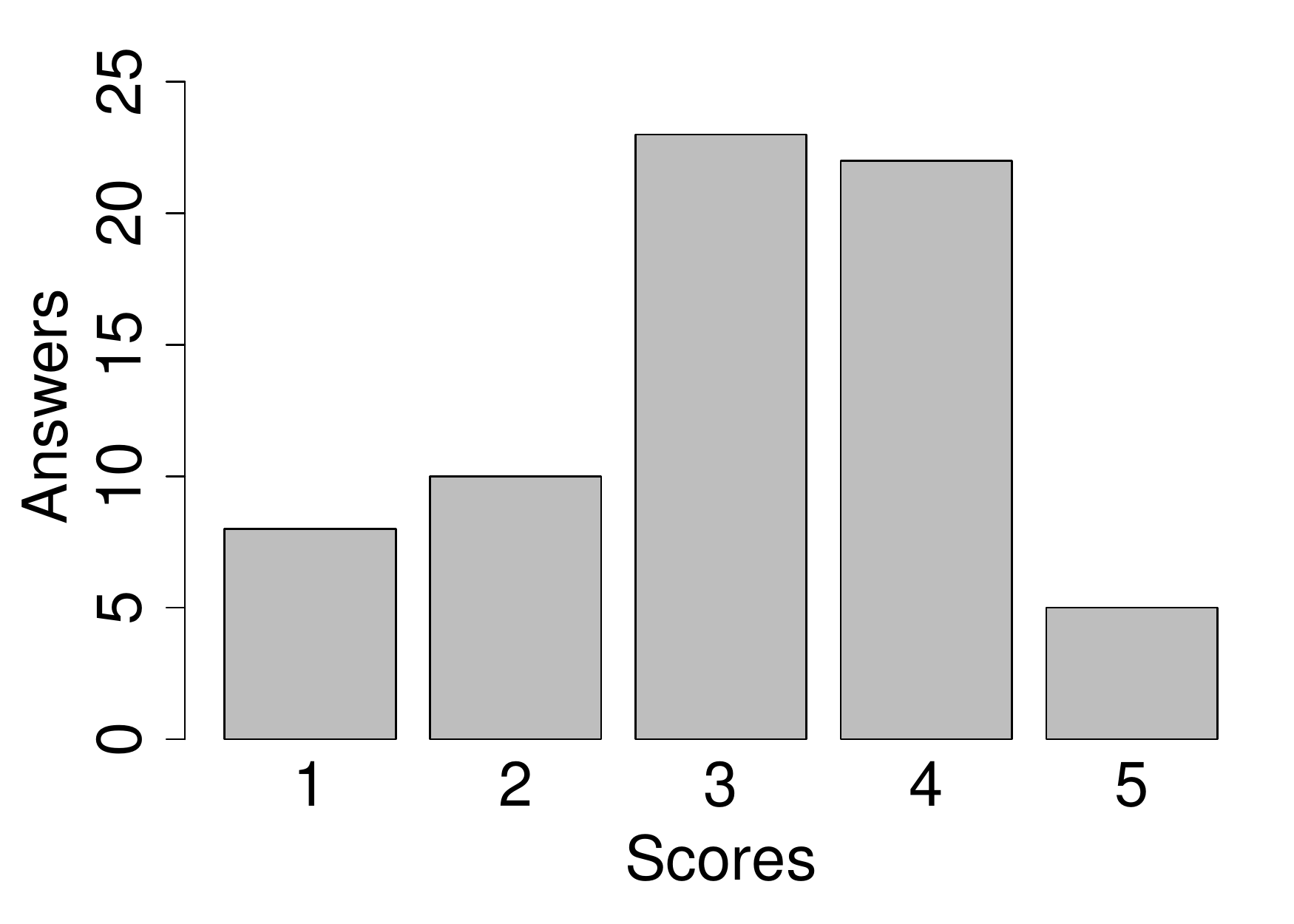}}
\quad
\subfloat[ref3][\socketio]{\includegraphics[width=0.3\textwidth]{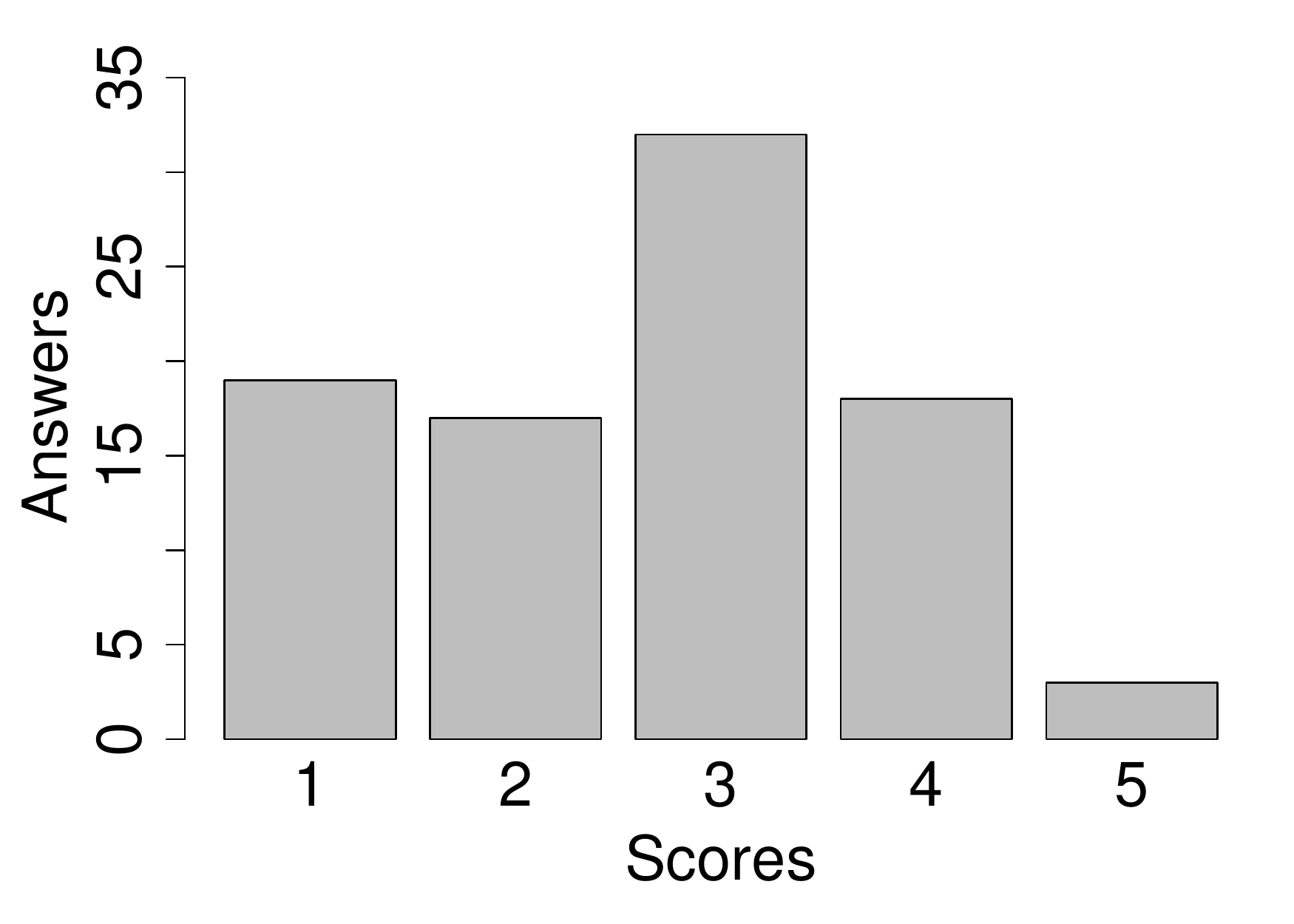}}
\quad
\caption{Survey answers}
\label{fig:survey-answers} 
\end{figure*} 
  
  
  The features are collected from client projects where the candidate experts contributed with at least one commit. In more detailed terms, suppose a candidate expert $c$; suppose also that $\mathit{Proj}_c$ are the projects where $c$ has made at least one commit (this set is provided by GitHub API). We iterate over $\mathit{Proj}_c$ to create a subset $\mathit{CliProj}_c$ containing only projects that depend on the target libraries. The features collected for $c$ are extracted from $\mathit{CliProj}_c$. 
After collecting this data, we found that 69\% of \sreact's candidate experts worked on a single client project; for \smongodb\ and \ssocketio, this percentage increases to 88\% and 87\%, respectively. By contrast, we found candidate experts working on 26 projects (\sreact), 5 projects (\smongodb) and 12 projects (\ssocketio).
    
\subsection{Ground Truth} 
\label{sec:ground-truth}

To create a ground truth with developers expertise on each target library, we conducted a survey with the candidate experts identified in Section~\ref{sec:experts}. For \sreact, which has 8,742 candidate experts, we sent the survey to a random sample of 2,185 developers (25\%). For \smongodb\ and \ssocketio, which have less candidates, we sent the survey to {\em all} candidate experts identified in Section~\ref{sec:experts}, i.e., to 454 and 608 developers, respectively. For each target library, we e-mailed the candidate experts, describing our research purpose and asking the following single question:\\[-0.3cm] 

\noindent {\em Could you please rank your expertise on [target library] in a scale from 1 (novice) to 5 (expert)?} \\[-0.3cm]
  
Table~\ref{tab:survey} summarizes the number of e-mails sent, the number of received answers, and the response ratio. The number of answers range from 68 (\smongodb) to 418 (\sreact) and the response ratio ranges from 15\% (\ssocketio\ and \smongodb) to 19\% (\sreact). 

\begin{table}[!t]
\centering
\caption{Survey Numbers}
\label{tab:survey}
\begin{tabular}{lrrr}
\toprule
{\bf Library} & {\bf Mails} & {\bf Answers} & {\bf Ratio}\\ 
\midrule
\react\ & 2,185 & 418 & 19\%\\
\mongodb\ & 454 & 68 & 15\%\\
\socketio\ & 608 & 89 & 15\%\\
\bottomrule
\end{tabular}
\end{table}

Figure~\ref{fig:survey-answers} shows the distribution of the survey answers.  For \sreact, 254 candidates (61\%) ranked themselves as experts in the library (scores 4--5); 110 candidates (26\%) declared an intermediate expertise (score 3), and 54 candidates (13\%) considered themselves as having a limited expertise (scores 1--2).  For \smongodb, the results are 40\% (experts), 34\% (intermediate expertise), and 26\% (limited expertise). For \ssocketio, the results are 24\%, 36\%, and 40\%, respectively.

\step{Ground Truth Limitations} The proposed ground truth is based on the developers'~perceptions about their expertise in the target libraries. Therefore, it is subjected to imprecisions and noise, since it is not realistic to assume the survey participants ranked themselves according to uniform and objective criteria. For example, some developers might have been more rigorous in judging their expertise, while others may have omitted their lack of experience on the studied libraries (see the Dunning-Kruger Effect~\cite{Kruger1999}). 
In order to try to reduce these issues, we made it clear to the participants that our interests were strictly academic and that we will never use their answers to commercial purposes. Finally, it is also worth mentioning that previous research has shown that 
self estimation is a reliable way to measure general programming experience, at least in a student population~\cite{Siegmund2014a}.

\subsection{Final Processing Steps}

We performed the following processing steps on the features collected for the developers that answered our survey.
  
\step{Missing Values} Missing values occur when it is not possible to compute a feature value. In our dataset, there are four features with missing values: {\em daysSinceFirstImport}, {\em daysSinceLastImport}, {\em daysBetweenImports}, and {\em avgDaysCommitsImportLibrary}. For these features, a missing value appears in candidate experts who have added an insufficient number of import statements to a client project (e.g., $\mathit{imports} = 0$). The percentage of candidate experts with missing values for these four features is relevant, as they appear in 45\% of the surveyed developers. To handle such cases, we replaced missing values at {\em daysSinceFirstImport} and {\em daysSinceLastImport} by a zero value, because candidate experts without import statements should not be viewed as long time library users. By contrast, missing values at {\em avgDaysCommitsImportLibrary} were replaced by the maximal observed value, because the respective candidate experts should have the highest values when compared to those who effectively added import statements. Finally, {\em daysBetweenImports} needs at least two imports to be calculated correctly. Therefore, we assigned a zero value when $\mathit{imports} = 1$, and $-1$ when $\mathit{imports} = 0$.\footnote{In fact, we tested different strategies for missing values, such as discarding all fields with missing values, applying different values, etc. However, the results never exceeded the ones based on the values proposed in this paragraph.}

\begin{figure}[!t]
  \centering
  \includegraphics[width=\columnwidth]{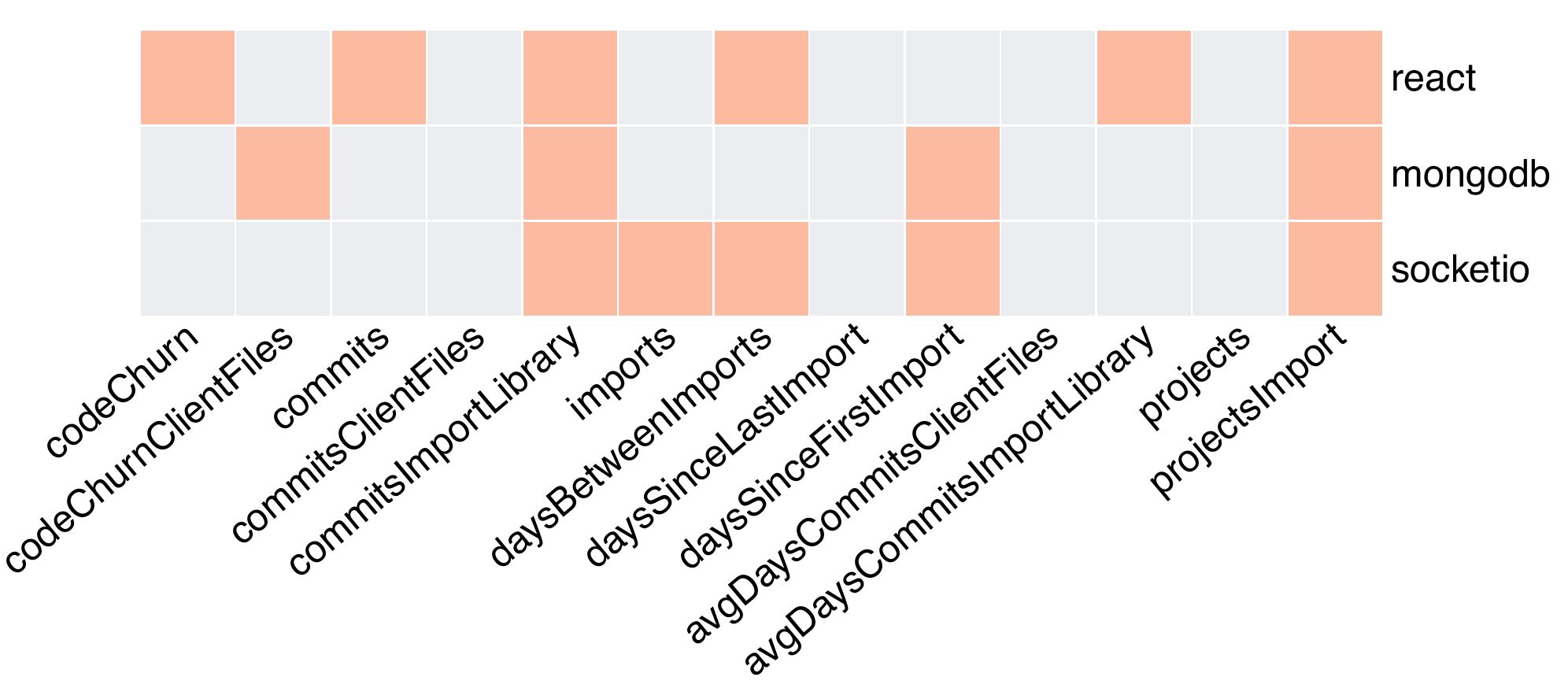}
  \caption{Correlation analysis; red cells are discarded due to high correlation.}
  \label{fig:correlations} 
\end{figure}

\step{Removing Correlated Features} Correlated features may contribute to inaccurate classifications due to their high association degree~\cite{Yu2003,chen2005}. To tackle this issue, we first used the {\em cor}\footnote{https://www.rdocumentation.org/packages/stats/versions/3.4.3/topics/cor} function from R's {\em stats} package to compute a matrix with Pearson coefficients for each pair of features. Then, we used the {\em findCorrelation}\footnote{https://www.rdocumentation.org/packages/caret/versions/6.0-79/topics/findCorrelation} function from R's {\em caret} package to identify pairs of features with a correlation greater than 0.7, as previously adopted in the literature~\cite{Bao2017}; in such cases, we measured the overall correlation of both features and discarded the highest one. Figure \ref{fig:correlations} shows a heatmap that summarizes this process. Red cells are features discarded due to a high correlation with another feature; gray cells denote features preserved by the correlation analysis, i.e.,~they are used in the classification process. As we can see, two features are correlated with at least one other feature, regardless the target library: {\em commitsImportLibrary} and {\em projectsImport}. As a result of this analysis, six, four, and five features were discarded at \sreact, \smongodb, and \ssocketio, respectively.

\step{Skewed Feature Values} Features with skewed distributions may impact the performance of machine learning classifiers~\cite{FeatureTransf,FeatureTransf2013}. 
We assume that skewed feature distributions are the ones where the mean---computed for the candidate experts included in the ground truth of a given target library---is at least four times greater than the median. By following this definition, four, six, and four features have a skewed behavior in \sreact, \smongodb, and \ssocketio, respectively. On the values of such features, we applied a $log$ transformation, as in another machine learning study~\cite{Boehm1999}.

\section{Methods}
\label{sec:methods}

In this section, we discuss the setup of the machine learning and clustering models, used on {\em RQ.1} and {\em RQ.2,} respectively.

\subsection{Machine Learning Setup and Algorithms}
\label{sec:machine_learning}

\newcommand{\sbar}[2]{{\color{darkgray}\rule{\dimexpr 0.6cm * #1 / 100}{5pt}\color{lightgray}\rule{\dimexpr 0.6cm * (100 - #1) / 100}{5pt}}}

\step{Number of Classes}
  Machine learning algorithms require a minimal number of samples on each class (or scores, in our terminology)~\cite{Raudys1991}. However, this condition is not followed by our data. For example,  for \sreact\ we collected expertise data about 418 developers, but only 24 developers (6\%) ranked themselves with score 2. 
  To attenuate this problem, we train and evaluate our models under two scenarios: (1) considering all five classes; (2) by transforming the data into the following ternary classification: \textit{novice} (scores 1--2), {intermediate} (score 3), and \textit{experts} (scores 4--5).
Furthermore, we only evaluate the scenario with five classes for \sreact.
The reason is because \smongodb\ and \ssocketio\ have fewer data points; for example, both libraries have classes with less than 10 samples.

 \step{Informed Over Sampling (SMOTE)} Besides having few samples for some classes, the ground truth is largely imbalanced, as illustrated in Figure \ref{fig:survey-answers}. For example, 87\% of the \sreact\ developers ranked themselves as having some knowledge on the framework (scores 3-5).
  It is well-known that machine learning classifiers tend to produce poor results when applied to imbalanced datasets~\cite{Japkowicz2002}.
To tackle this problem,
  we used a technique called 
  Informed Over Sampling (SMOTE)~\cite{Chawla2002}, which balances a dataset by producing and inserting synthetic but similar observations to minority classes (but only in the training part of the dataset). 
SMOTE was previously used in machine learning approaches to several software engineering problems, including defect prediction~\cite{Tan2015}, mobile apps analysis~\cite{Li2016}, self-admitted technical debt detection~\cite{Zampetti2017}, and identification of security issues from commit messages and bug reports~\cite{Zhou2017}. 
In our problem, we used SMOTE over the minority class, on both scenarios.  SMOTE has two parameters: number of the nearest neighbours (KNN) and the percentage of synthetic instances to create. After some initial tests, we set up these parameters to 3 and 30\%, respectively. This setup results in a minority class increased by 30\%; and the new data points are synthesized by considering 3-nearest neighbours of the existing ones (KNN parameter).

\step {Machine Learning Classifiers} We evaluate two well-known machine learning classifiers: Random Forest~\cite{Breiman2001} and SVM~\cite{Weston1998}. We compare the results of these classifiers with a ZeroR baseline, which simply predicts the majority class, ignoring all feature values. We do not compare with previous expertise identification approaches (e.g., ~\cite{Mockus2002b,Fritz2007,Fritz2010,Fritz2014,Schuler2008}) because they are not proposed to measure expertise on libraries and frameworks, but on internal elements of a software project. We use $k$-fold stratified cross-validation to evaluate the results of these classifiers. Stratified cross-validation is a variant of $k$-fold cross-validation where folds contain approximately the same proportion of each class.  We set $k$ to 5, to avoid testing models in small folds, particularly in small classes, as occur in \smongodb\ and \ssocketio. Another important step is the tuning of the classifiers parameters. We rely on a grid search strategy for hyper-parameters with cross validation to find the best parameters settings for each classifier~\cite{gridSearch}. 

 \setcounter{table}{6} 
\begin{table*}[!t]
\centering
\caption{Results for 3 classes: novice (scores 1-2), intermediate (score 3), and expert (scores 4-5)}
  \label{tab:results3}
\begin{tabular}{lrrrrrrrrr}
\toprule
  & \multicolumn{3}{c}{\bf \react} & \multicolumn{3}{c}{\bf \mongodb}    & \multicolumn{3}{c}{\bf \socketio} \\
                & {\bf RForest}    & {\bf SVM}   & {\bf Baseline}   & {\bf RForest} & {\bf SVM} & {\bf Baseline} & {\bf RForest}  & {\bf SVM} & {\bf Baseline} \\ \midrule
      {\bf Kappa}                    &  0.09 \sbar{09}{}  & 0.03 \sbar{03}{} &  0.00 \sbar{00}{} &  0.35 \sbar{35}{}  &   0.25 \sbar{25}{} &  0.00 \sbar{00}{}  &   0.16 \sbar{16}{}  &  0.25 \sbar{25}{} &  0.00 \sbar{00}{}  \\
      {\bf AUC}                      &  0.56 \sbar{56}{}  & 0.51 \sbar{51}{} &  0.50 \sbar{50}{} &  0.70 \sbar{70}{}  &   0.56 \sbar{56}{} &  0.50 \sbar{50}{}  &   0.60 \sbar{60}{}  &  0.71 \sbar{71}{} &  0.50 \sbar{50}{}  \\      
      {\bf Precision (Novice)}       &  0.14 \sbar{14}{}  & 0.60 \sbar{60}{} &  0.00 \sbar{00}{} &  0.50 \sbar{50}{}  &   0.47 \sbar{47}{} &  0.00 \sbar{00}{}  &   0.52 \sbar{52}{}  &  0.54 \sbar{54}{} &  0.40 \sbar{40}{}  \\
      {\bf Precision (Intermediate)} &  0.34 \sbar{34}{}  & 0.00 \sbar{00}{} &  0.00 \sbar{00}{} &  0.62 \sbar{62}{}  &   0.17 \sbar{17}{} &  0.00 \sbar{00}{}  &   0.29 \sbar{29}{}  &  0.59 \sbar{59}{} &  0.00 \sbar{00}{}  \\
      {\bf Precision (Expert)}       &  0.65 \sbar{65}{}  & 0.61 \sbar{61}{} &  0.61 \sbar{61}{} &  0.55 \sbar{55}{}  &   0.57 \sbar{57}{} &  0.40 \sbar{40}{}  &   0.43 \sbar{43}{}  &  0.48 \sbar{48}{} &  0.00 \sbar{00}{}  \\
      {\bf Recall (Novice)}          &  0.09 \sbar{09}{}  & 0.06 \sbar{06}{} &  0.00 \sbar{00}{} &  0.50 \sbar{50}{}  &   0.68 \sbar{68}{} &  0.00 \sbar{00}{}  &   0.61 \sbar{61}{}  &  0.78 \sbar{78}{} &  1.00 \sbar{100}{} \\
      {\bf Recall (Intermediate)}    &  0.18 \sbar{18}{}  & 0.00 \sbar{00}{} &  0.00 \sbar{00}{} &  0.57 \sbar{57}{}  &   0.09 \sbar{09}{} &  0.00 \sbar{00}{}  &   0.19 \sbar{19}{}  &  0.19 \sbar{19}{} &  0.00 \sbar{00}{}  \\
      {\bf Recall (Expert)}          &  0.83 \sbar{83}{}  & 1.00 \sbar{100}{} & 1.00 \sbar{100}{} &  0.63 \sbar{63}{}  &   0.75 \sbar{75}{} & 1.00 \sbar{100}{}  &   0.56 \sbar{56}{}  &  0.56 \sbar{56}{} &  0.00 \sbar{00}{}  \\
      {\bf F-measure}                &  0.36 \sbar{36}{}  & 0.29 \sbar{29}{} &  0.25 \sbar{25}{} &  0.56 \sbar{56}{}  &   0.44 \sbar{44}{} &  0.19 \sbar{19}{}  &   0.42 \sbar{42}{}  &  0.46 \sbar{46}{} &  0.19 \sbar{19}{}  \\
\bottomrule
\end{tabular}
\end{table*}

\step{Evaluation Metrics}
We evaluate the classifiers using precision, recall, F-measure, and AUC (Area Under the Receiver Operating Characteristic Curve).  To compute AUC, we use an implementation recommended for multi-class classifications. This implementation is provided as an R package by Microsoft Azure's data science team.\footnote{https://github.com/Azure/Azure-MachineLearning-DataScience} Further, to compute F-measure, we first compute the average precision and recall, considering all classes. The reported F-measure is the harmonic mean of the average precision and average recall. We also report Cohen's kappa, which is also a measure of classifier performance, particularly useful on imbalanced datasets~\cite{kappa}.

\subsection{Clustering Setup and Algorithm}
\label{sec:clustering_setup}

We use clustering to investigate more closely the relation of feature values and library expertise ({\em RQ.2}). To this purpose, we use $k$-means, which is a widely popular clustering algorithm. In software engineering, $k$-means was used to support many tasks, including detecting mobile apps with anomalous behavior~\cite{Gorla2014}, test case prioritization~\cite{Arafeen2013}, and to characterize build failures~\cite{Vassallo2017}. A key challenge when using $k$-means is to define the appropriate number $k$ of clusters. There are methods proposed to help on this task, such as the elbow~\cite{Ng2000} and silhouette methods~\cite{Rousseeuw1987}. However, they also depend on interpretation and subjective decisions~\cite{Ng2000}. For this reason, we follow an alternative procedure, as described next. We execute $k$-means multiple times, starting with $k=2$ and incrementing it after each execution. For each $k$, we analyze the resulting clusters, searching for clusters dominated by experts. For \sreact, we search for clusters with at least 70\% of experts (since \sreact\ has a higher percentage of experts in the ground truth, close to 61\%); for \smongodb\ and \ssocketio---which have less experts,~40\% and 24\%, respectively---we search for clusters with at least 60\% of experts. We stop after finding at least one cluster attending the proposed thresholds. Table \ref{tab:kmeans} shows data on each  execution; for each $k$, it shows the percentage of experts of the cluster with the highest percentage of experts. For \sreact, we select 3 clusters, since it leads to a cluster with 74\% of experts. For \smongodb, we also select 3 clusters, including a cluster with 65\% of experts. For \ssocketio, there are 5 clusters and one has 75\% of experts. 

 \setcounter{table}{4} 
\begin{table}[!ht]
  \centering
  \caption{Cluster with the highest percentage of experts (values in bold define the selected number of clusters)}
  \label{tab:kmeans}
  \begin{tabular}{lrrrr}
  \toprule
  \multirow{2}{*}{\bf Library}    & \multicolumn{4}{c}{\bf $k$} \\
             & 2     & 3     & 4    & 5    \\ \midrule
  \sreact    & 66  & {\bf 74}  & -    & -    \\
  \smongodb  & 57  & {\bf 65}  & -    & -    \\
  \ssocketio & 39  & 44  & 44 & {\bf 75} \\
  \bottomrule
  \end{tabular}
\end{table}

 \setcounter{table}{5} 
\begin{table}[!t]
\centering
\caption{Machine learning results for 5 classes (\react)}
\label{tab:results5}
\begin{tabular}{lrrr}
      \toprule
      & {\bf RForest} & {\bf SVM}    & {\bf Baseline} \\ \midrule
      {\bf Kappa}                     &    0.09 \sbar{09}{} &   0.05 \sbar{05}{}    &   0.00 \sbar{00}{} \\
      {\bf AUC}                       &    0.52 \sbar{52}{} &   0.53 \sbar{53}{}    &   0.50 \sbar{50}{} \\      
      {\bf Precision (Novice 1)}      &    0.25 \sbar{25}{} &   0.00 \sbar{00}{}    &   0.00 \sbar{00}{} \\
      {\bf Precision (Novice 2)}      &    0.07 \sbar{07}{} &   0.00 \sbar{00}{}    &   0.00 \sbar{00}{} \\
      {\bf Precision (Intermediate)}  &    0.35 \sbar{35}{} &   0.23 \sbar{23}{}    &   0.00 \sbar{00}{} \\
      {\bf Precision (Expert 4)}      &    0.50 \sbar{50}{} &   0.48 \sbar{48}{}    &   0.46 \sbar{46}{} \\
      {\bf Precision (Expert 5)}      &    0.29 \sbar{29}{} &   0.00 \sbar{00}{}    &   0.00 \sbar{00}{} \\
      {\bf Recall (Novice 1)}         &    0.07 \sbar{07}{} &   0.00 \sbar{00}{}    &   0.00 \sbar{00}{} \\
      {\bf Recall (Novice 2)}         &    0.04 \sbar{04}{} &   0.00 \sbar{00}{}    &   0.00 \sbar{00}{} \\
      {\bf Recall (Intermediate)}     &    0.27 \sbar{27}{} &   0.10 \sbar{10}{}    &   0.00 \sbar{00}{} \\
      {\bf Recall (Expert 4)}         &    0.77 \sbar{77}{} &   0.98 \sbar{98}{}    &  1.00 \sbar{100}{} \\
      {\bf Recall (Expert 5)}         &    0.10 \sbar{10}{} &   0.00 \sbar{00}{}    &   0.00 \sbar{00}{} \\
      {\bf F-measure}                 &    0.24 \sbar{24}{} &   0.15 \sbar{15}{}    &   0.13 \sbar{13}{} \\
\bottomrule
\end{tabular}
\end{table}
 \setcounter{table}{7}

\section{Results}
\label{sec:results}


\noindent{\em (RQ.1) How accurate are machine learning classifiers when used to identify library experts?} \\[-0.25cm]

Table~\ref{tab:results5} presents the results of the machine learning classifiers for five classes. The results are provided only for \sreact, since \smongodb\ and \ssocketio\ do not have sufficient samples to perform a classification using five classes, as explained in Section~\ref{sec:machine_learning}. For almost all performance metrics and classifiers, the results are not good. For example, kappa is 0.09 and AUC is 0.56 for Random Forest. Precision ranges from 0.00 (Novice 2, SVM) to 0.50 (Expert 4, Random Forest). F-measure is 0.24 (Random Forest) and 0.15 (SVM), against 0.13 with the ZeroR baseline.

Table~\ref{tab:results3} presents the results for three classes (scores 1-2, score 3, scores 4-5). First, we discuss the results of Random Forest. For this classifier, kappa varies from 0.09 (\sreact) to 0.35 (\smongodb); AUC ranges from 0.56 (\sreact) to 0.70 (\smongodb). Precision results are greater for experts than for novices, both for \sreact\ (0.65 vs 0.14) and \smongodb\ (0.61 vs 0.60), while \ssocketio\ has the highest precision for novices (0.52). Recall ranges from 0.09 (\sreact, novices) to 0.83 (\sreact, experts). F-measure is 0.36 (\sreact), 0.56 (\smongodb), and 0.42 (\ssocketio). By contrast, the baseline results for F-measure are 0.25 (\sreact) and 0.19 (\smongodb\ and \ssocketio). In the same scenario, SVM results are in 13 out of 27 combinations of metrics and libraries lower than the ones of Random Forest; they are also just slightly greater than ZeroR. 


\begin{formal}
For five classes, machine learning classifiers have a maximal F-measure of 0.24 (\sreact). For three classes, F-measure reaches 0.56 (\smongodb) and precision on identifying experts reaches 0.65 (\sreact, experts).
\end{formal}

\noindent{\em (RQ.2) Which features best distinguish library experts?} \\[-0.25cm]

First, Table~\ref{tab:clustering_result} shows the percentage of novices (scores 1-2), intermediate (score 3), and experts (scores 4-5) in the clusters of each library. The table also shows the number of developers in each cluster.
As defined in Section \ref{sec:clustering_setup}, for \sreact\ and \smongodb, we have 3 clusters; for \ssocketio, we have 5 clusters. In Table~\ref{tab:clustering_result}, the clusters are sorted by percentage of experts. Therefore, Cluster 1 is the experts' cluster in each library. In \sreact, 74\% of the developers in this cluster ranked themselves as experts and only 3\% as novices. For \smongodb\ and \ssocketio, Cluster 1 includes 65\% and 75\% of experts, respectively. By contrast, it has only 12\% and 0\% of novices, respectively.   
The number of developers in the experts' cluster ranges from 4 (\ssocketio) to 97 developers (\sreact). However, the ground truth has also more \sreact\ experts (254 vs 21 developers, respectively). Interestingly, in \ssocketio, Cluster 5 should be viewed as a novice's clusters; 67\% of its members are novices and the cluster does not include any expert.

\begin{formal}
In the three studied libraries, there are clusters dominated by experts. These clusters have 74\% (\sreact), 65\% (\smongodb), and 75\% (\ssocketio) of experts.
\end{formal}

\newcommand{\lbar}[1]{{\color{darkgray}\rule{\dimexpr 0.75cm * #1 / 100}{5pt}\color{lightgray}\rule{\dimexpr 0.75cm * (100 - #1) / 100}{5pt}}}

\begin{table}[!t]
  \centering
  \caption{Clustering results (cluster 1 has the highest \% of experts)}
  \label{tab:clustering_result}
  \begin{tabular}{lrrrr}
    \toprule
    {\bf Cluster}  & {\bf \% Novices} & {\bf \% Intermediate} & {\bf \% Experts} & {\bf \# Devs}  \\ \midrule
    \multicolumn{5}{l}{\react} \vspace{3pt} \\
    {\bf C1}   & 0.03 \lbar{03}    & 0.23 \lbar{23}         & 0.74 \lbar{74}    & 97  \\
    C2   & 0.12 \lbar{12}    & 0.28 \lbar{28}         & 0.60 \lbar{60}    & 129 \\ 
    C3   & 0.18 \lbar{18}    & 0.27 \lbar{27}         & 0.55 \lbar{55}    & 192 \\ \midrule
    \multicolumn{5}{l}{\mongodb} \vspace{3pt} \\
    {\bf C1}   & 0.12 \lbar{12}    & 0.24 \lbar{24}         & 0.65 \lbar{65}    & 17  \\
    C2   & 0.21 \lbar{21}    & 0.43 \lbar{43}         & 0.36 \lbar{36}    & 14  \\ 
    C3   & 0.35 \lbar{35}    & 0.35 \lbar{35}         & 0.30 \lbar{30}    & 37  \\ \midrule
    \multicolumn{5}{l}{\socketio} \vspace{3pt} \\
    {\bf C1}   & 0.00 \lbar{00}    & 0.25 \lbar{25}         & 0.75 \lbar{75}    & 4   \\
    C2   & 0.29 \lbar{29}    & 0.36 \lbar{36}         & 0.36 \lbar{36}    & 28  \\
    C3   & 0.33 \lbar{33}    & 0.33 \lbar{33}         & 0.33 \lbar{33}    & 15  \\
    C4   & 0.50 \lbar{50}    & 0.40 \lbar{40}         & 0.10 \lbar{10}    & 30  \\
    C5   & 0.67 \lbar{67}    & 0.33 \lbar{33}         & 0.00 \lbar{00}    & 12  \\
    \bottomrule
  \end{tabular}
\end{table}

\begin{figure*}[!t]
\centering
\subfloat[ref1][\react]{\includegraphics[width=0.32\textwidth]{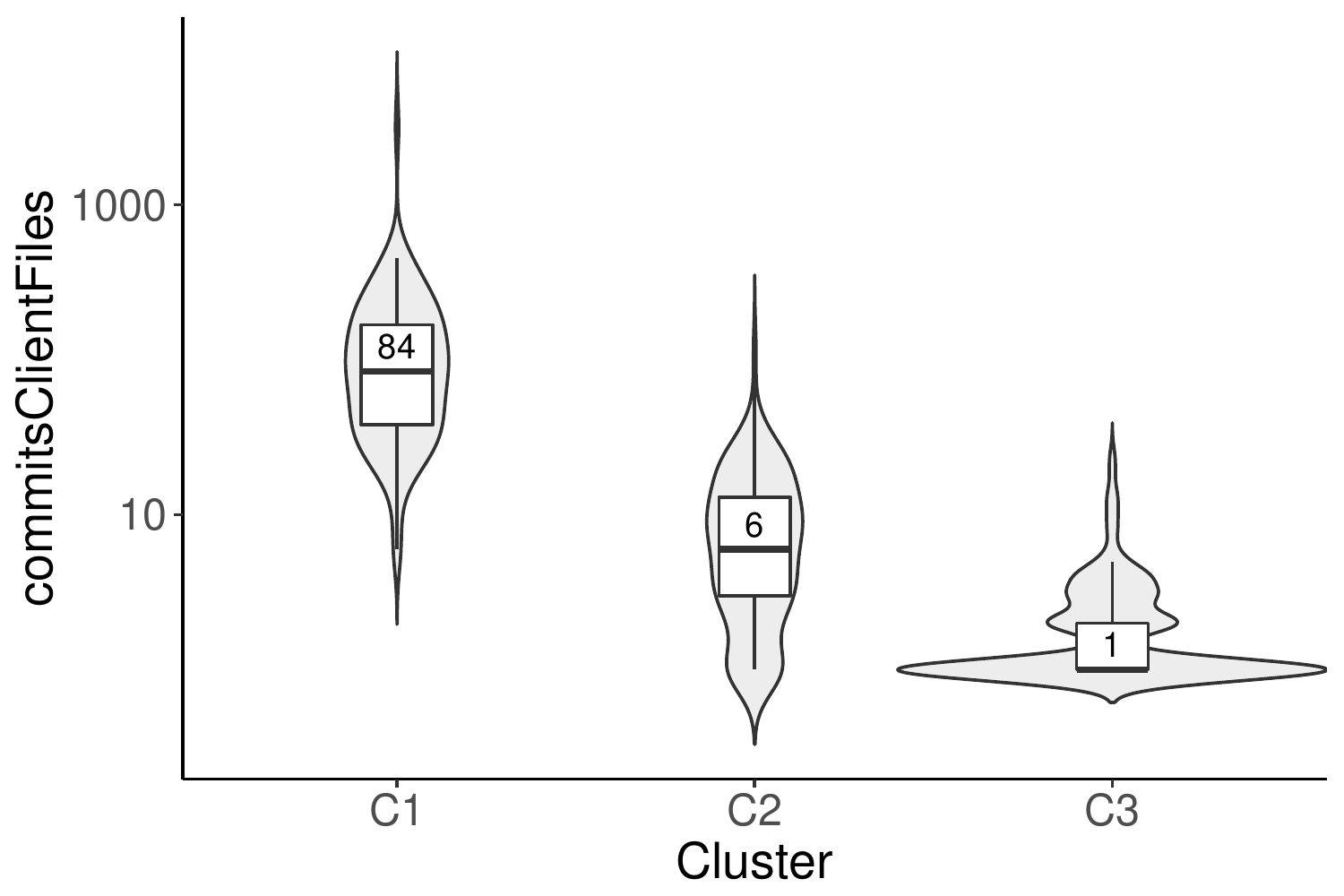}}
\quad
\subfloat[ref2][\mongodb]{\includegraphics[width=0.32\textwidth]{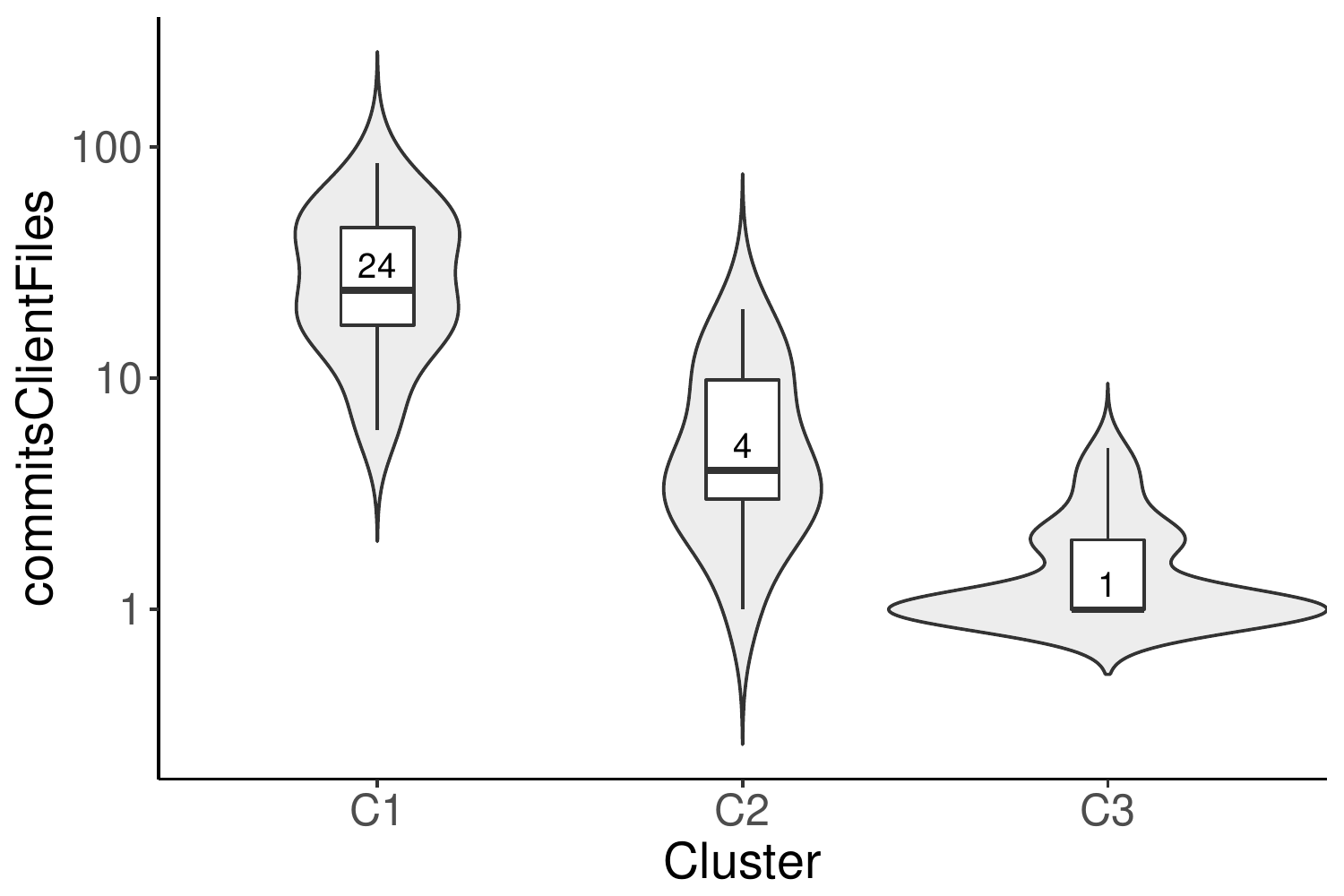}}
\quad
\subfloat[ref3][\socketio]{\includegraphics[width=0.32\textwidth]{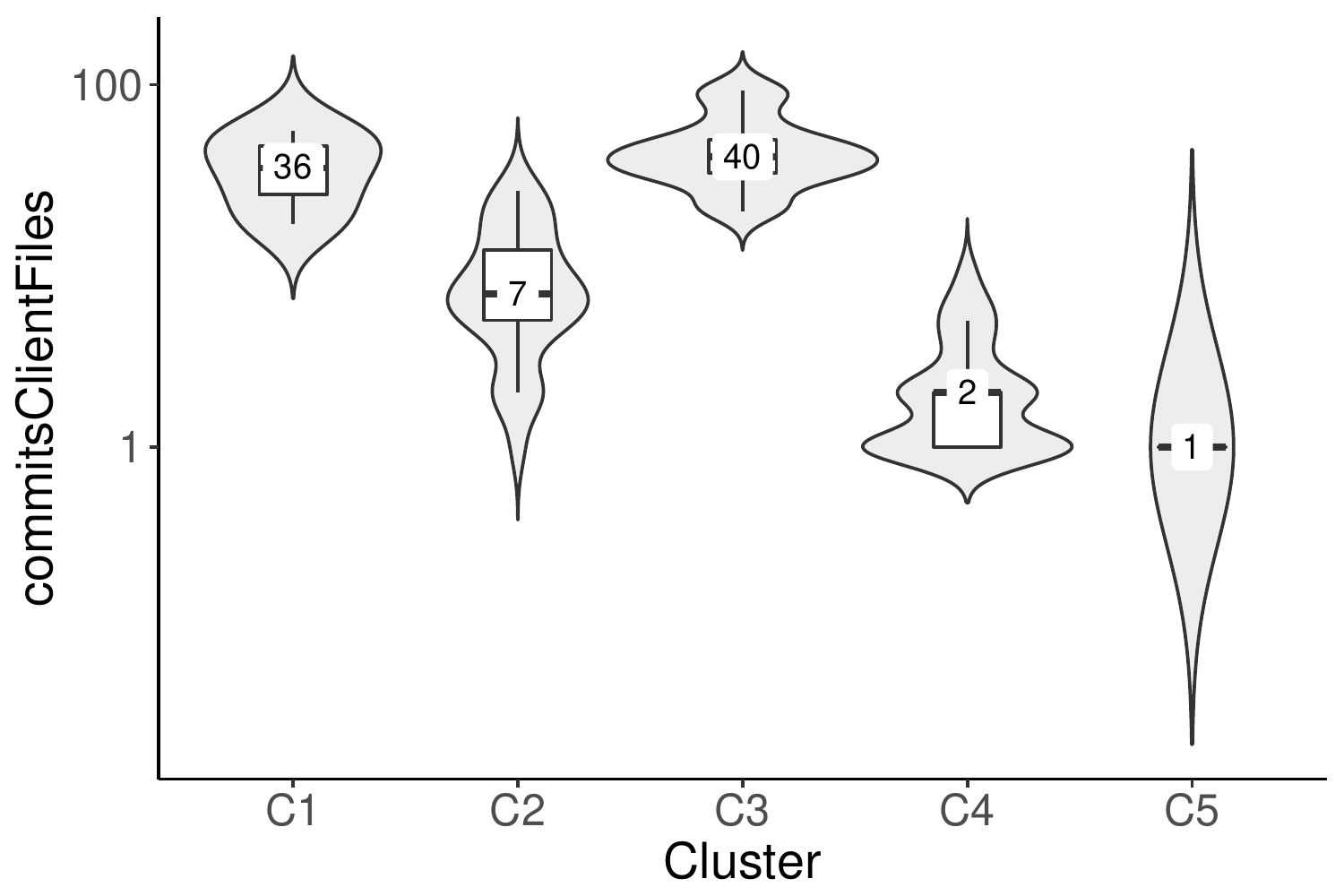}}
\quad
\caption{Distributions of {\em commitsClientFiles} values for each cluster/library. Cluster 1 (experts) has higher values than other clusters, except for \ssocketio.}
\label{fig:commitsClientFiles} 
\end{figure*}

We also compare the distributions of feature values, for the developers in each cluster.  For each feature $F$, we compare $F$'s distribution in Cluster 1 (experts) with the cluster whose median of $F$'s distribution is closest to the one of Cluster 1. In other words, this cluster tends to be the most similar to Cluster 1, among the remaining clusters; our goal is to assess the magnitude (effect size) and direction of this similarity. First, we use a Mann-Whitney test to confirm that the distributions of $F$'s values in both clusters are statistically distinct, assuming a $p$-value of 0.05. Furthermore, and more interestingly, we measure the magnitude and direction of the difference, using Cliff’s delta. As in other works~\cite{Grissom2005, Romano2006, Linares2013, Tian2015}, we interpret Cliff's delta as negligible for $d < 0.147$, small for $0.147 \leq d < 0.33$, medium for $0.33 \leq d < 0.474$, and large for $d \geq 0.474$.

\begin{table}[!t]
  \centering
  \caption{Comparing feature distributions using Cliff's delta: Experts vs Cluster with the closest median ($\circ$ means similar distributions, according to Mann-Whitney, $p$-value= 0.05)}
  \label{tab:cliffdelta}
  \begin{tabular}{lcc} \toprule
  {\bf Feature} & {\bf Effect size} & {\bf Relationship} \\ \midrule
  \multicolumn{3}{l}{\react} \vspace{3pt}        \\
  codeChurnClientFiles        & large        & $+$  \\
  commitsClientFiles          & large        & $+$  \\
  imports                     & large        & $+$  \\
  daysSinceLastImport         & large        & $+$  \\
  daysSinceFirstImport        & medium       & $+$  \\
  avgDaysCommitsClientFiles   & large        & $-$  \\
  projects                    & large        & $+$  \\ \midrule
  \multicolumn{3}{l}{\mongodb} \vspace{3pt}     \\
  codeChurn                   & large        & $+$  \\
  commits                     & large        & $+$  \\
  commitsClientFiles          & large        & $+$  \\
  imports                     & large        & $+$  \\
  daysBetweenImports          & large        & $+$  \\
  daysSinceLastImport         & medium       & $+$  \\
  avgDaysCommitsClientFiles   & large        & $-$  \\
  avgDaysCommitsImportLibrary & large        & $-$  \\
  projects                    & large        & $+$  \\ \midrule
  \multicolumn{3}{l}{\socketio} \vspace{3pt}    \\
  codeChurn                   & $\circ$     & $\circ$ \\
  codeChurnClientFiles        & $\circ$     & $\circ$ \\
  commits                     & $\circ$     & $\circ$ \\
  commitsClientFiles          & $\circ$     & $\circ$ \\
  daysSinceLastImport         & $\circ$     & $\circ$ \\
  avgDaysCommitsClientFiles   & $\circ$     & $\circ$ \\
  avgDaysCommitsImportLibrary & $\circ$     & $\circ$ \\ 
  projects                    & large        & $+$  \\
  \bottomrule
  \end{tabular}
\end{table}

\begin{figure}
  \centering
  \includegraphics[width=0.39\textwidth]{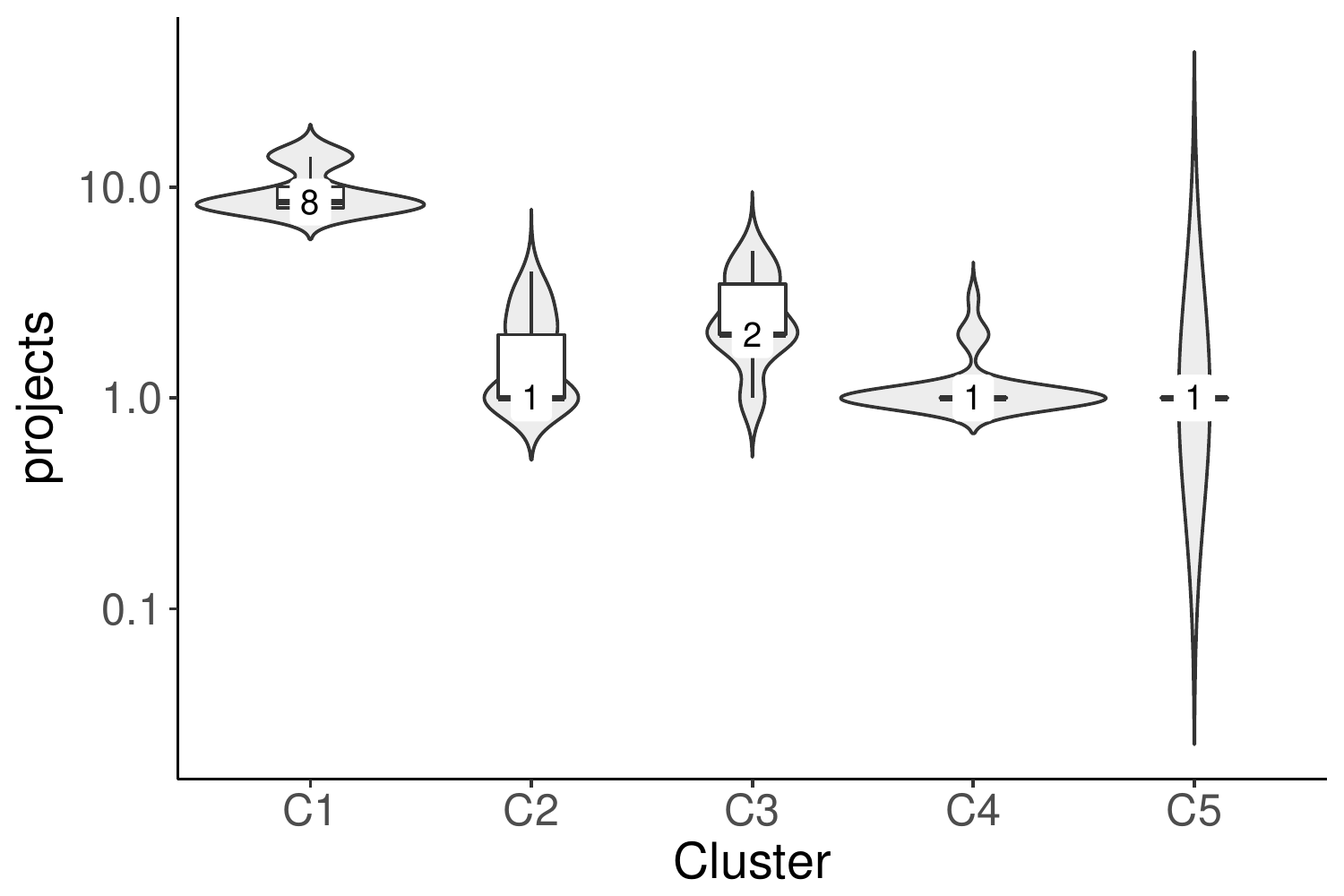}
  \caption{Distributions of {\em projects} values for \ssocketio\ clusters. Cluster 1 (experts) has higher values than other clusters.}
  \label{fig:boxplot_projects_socketio}
\end{figure}

Table~\ref{tab:cliffdelta} shows the results. For \sreact, there is a {\em large} difference for the distributions of all features in Cluster 1, with the exception of {\em daysSinceFirstImport}, which has a {\em medium} effect size.  The direction is mostly positive ($+$), i.e.,  developers in Cluster 1 have higher feature values than the ones in the second most similar cluster (in summary, they are more active on client files). The exception regards the distributions of {\em avgDaysCommitsClientFiles}, i.e., experts tend to commit more frequently to \sreact\ client files---in lower time intervals---than developers of the second cluster. In general, the results for \smongodb\ follow the same patterns observed for \sreact; the main exception is that a {\em medium} difference is observed for {\em daysSinceLastImport}. However, in the case of \socketio\, there is a major change in the statistical tests. First,  Cliff's delta reports a {\em large} difference for a single feature: number of projects the developers have committed to ({\em projects}). According to Mann-Whitney tests, the remaining feature distributions are statistically indistinct. To visually illustrate these results, Figure~\ref{fig:commitsClientFiles} shows violin plots with the distribution on each cluster of {\em commitsClientFiles}, for the three studied libraries. We can see a {\em large} difference between the distributions of Cluster 1 and Cluster 2, both for \sreact\ and \smongodb. By contrast, for \ssocketio, there is no clear difference between the distributions of Cluster 1 and Cluster 3 (cluster with the median closest to Cluster 1). Finally, Figure \ref{fig:boxplot_projects_socketio} shows boxplots with {\em projects} distribution for \ssocketio. In this case, we can see a clear difference between Cluster 1 (1st quartile is 8 projects; median is 8.5 projects) and Cluster 3 (1st quartile is one project; median is two projects).  

\begin{formal}
For \sreact\ and \smongodb, 
developers in the experts cluster are more active on GitHub than developers in other clusters, regarding most features. However, for \ssocketio, experts are only distinguished by the number of projects they worked on.
\end{formal}

To conclude, it is important to mention that the feature values are different for experts in each library. For example, experts in \react\ (Cluster 1) perform 84 commits at client files, against 24 commits for \smongodb's experts (median values, see Figure \ref{fig:commitsClientFiles}). Our hypothesis is that \sreact\ is a more complex framework than \smongodb, besides targeting a different domain. As a result, it is no trivial to define feature thresholds to classify experts; furthermore, these thresholds should not be reused across libraries.

 \setcounter{figure}{5}
\begin{figure*}[!t]
  \centering
  \includegraphics[width=\textwidth]{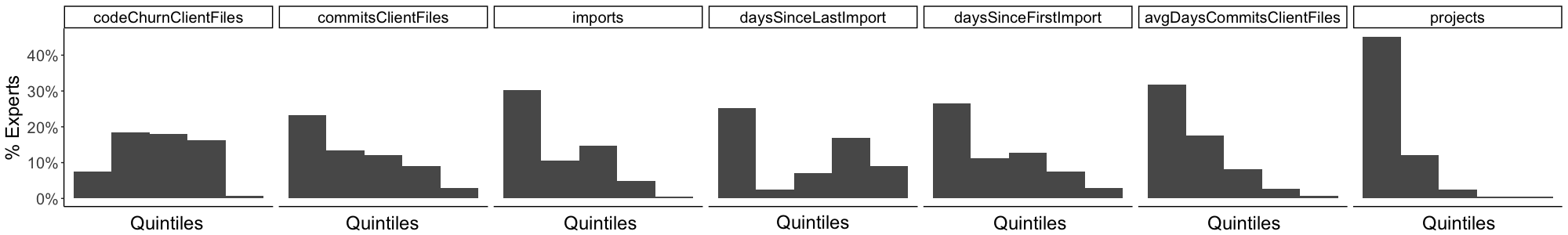}
  \caption{Percentage of \sreact\ experts  by quintiles of feature distributions. For most features, there is an important proportion of experts in lower quintiles.}
  \label{fig:histograms_react} 
\end{figure*}  
\setcounter{figure}{4}

\section{Discussion and Practical Usage}
\label{sec:discussion}


\subsection{Relevance and Key Findings}

In the survey to create the ground truth, we only asked for a score (in a 5-point scale). Despite that, we received some comments about the relevance of approaches to predict developers expertise in specific programming technologies, as in the following answers:\\[-0.35cm]

\noindent{\em What you are doing sounds very interesting and worthwhile to the developer's community at large.} (P021)\\[-0.35cm]

\noindent{\em Technical recruiting seems to be an extremely valid use-case for accurately assess the skills of devs based on their GitHub contributions, which could  lead to a profitable product.
} (P183)\\[-0.35cm]

We associate the high number of responses received in the survey (575 answers) to the 
relevance and potential practical value of the problem we proposed to investigate, which was rapidly viewed in this way by the surveyed GitHub users.

As mentioned in one of the previous answers, the main interest of companies is on accurately identifying experts in a given programming technology. In this particular context, precision is more important than recall, since companies do not need to identify all skilled engineers in a given technology, but only a few of them.
When approaching the problem using machine learning classifiers, we achieved a maximal precision of 65\% for the experts class
(scores 4-5, Random Forest, \sreact). 
In the same scenario, the baseline precision is 0.61. Therefore, this result casts doubts on the practical value of using machine learning in this problem. 
By contrast, when  using unsupervised techniques, based on clustering ($k$-means), we were able to identify clusters with 74\% (\sreact), 65\% (\smongodb), and 75\% (\ssocketio) of experts. If we consider that predicting expertise on programming technologies is 
a relevant but challenging problem, we claim that precision values close to 70\%---across multiple libraries---can sustain the practical adoption of automatic classifiers based on features extracted from GitHub activity. 
Even so, unsupervised techniques should be carefully used, as their gains may vary according to the library (see \sreact\ clusters).
It is also worth mentioning that such classifiers do not replace but complement traditional mechanisms for assessing developers expertise, like interviews and curriculum analysis. 

\subsection{Practical Usage}  
\label{sec:practicalusage}

Suppose a library $\mathcal{L}$ with developers grouped in clusters $\mathcal{C}_1, \ldots, \mathcal{C}_n$, after following the methodology proposed in this paper. Suppose that $\mathcal{C}_1$ groups the experts in $\mathcal{L}$. Given these clusters, suppose we want to assess the expertise of a new developer $d$ on $\mathcal{L}$, e.g., we are part of a company that heavily depends on $\mathcal{L}$ and we want to assess the expertise of $d$ in this library, before hiring her. In this case, we should retrieve the feature vector $\mathcal{F}_d$ for $d$, based on her activities on GitHub. Then, we compute the Euclidean distance between $\mathcal{F}_d$ and the centroid of each cluster $\mathcal{C}_i$, for $i=1, \ldots, n$. If the smallest distance is found between $\mathcal{F}_d$ and $C_1$'s centroid, we can assume that $d$ is more similar to the experts in $\mathcal{L}$ and therefore she has high chances of also being an expert in this library. Otherwise, our method fails to predict $d$'s expertise in $\mathcal{L}$, i.e., she can be or not an expert. It is also straightforward to identify expertise in multiple libraries. In this case, we only need to compute the intersection of experts in each library.

\subsection{Triangulation with Linkedin Profiles}

\begin{figure}[!t]
  \centering
  \includegraphics[width=\columnwidth]{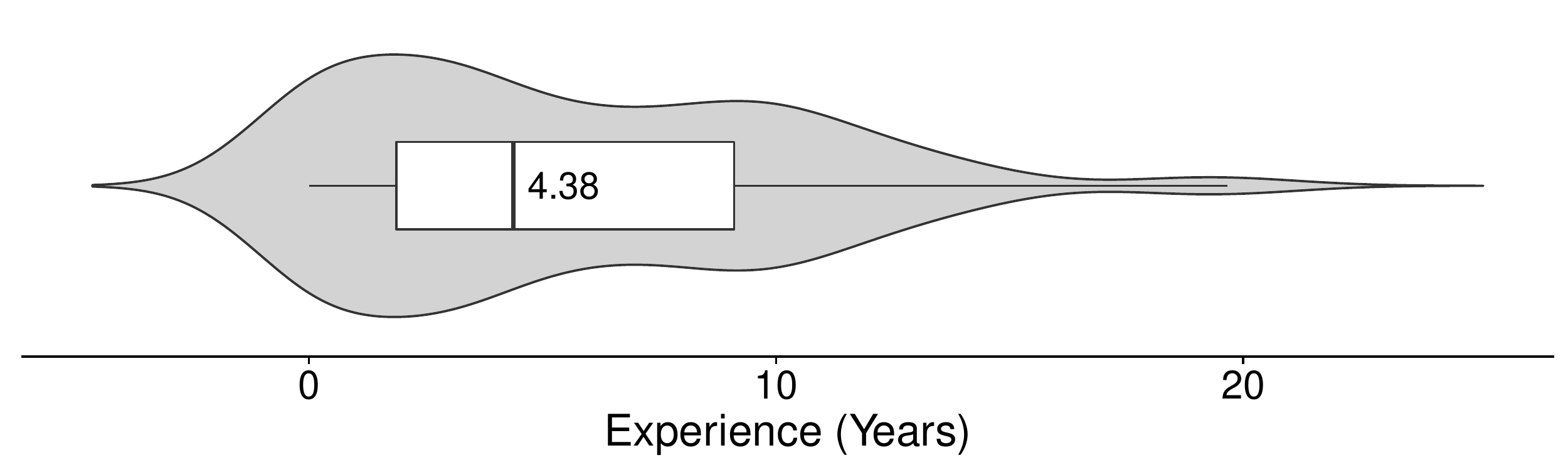}
  \caption{Years of experience on \sreact\ of developers predicted as experts}
  \label{fig:violin_experience} 
\end{figure}

To provide preliminary evidence on the value of the procedure described in the previous section to identify experts, we triangulated its results with expertise information available on Linkedin, starting with \sreact\ experts. First, we mapped each \sreact\ developer who did not answer our survey---and therefore was not considered at all in {\em RQ.1} and {\em RQ.2}---to one of the clusters produced for \sreact, as discussed before. 263 (out of 2,129 developers, 12\%) were mapped to the experts cluster. After that, the first author of this paper manually searched for the Linkedin page of these developers, looking for their names and possibly e-mails on Linkedin (when available, he also compared the profile photos, at Linkedin and GitHub). He was able to find the Linkedin profile of 160 developers (61\%). Finally, he manually examined these profiles, searching for evidences of expertise on \sreact. 115 developers (72\%) explicitly refer to \sreact\ on their Linkedin short bios, on the description of the projects they worked on, or in the list of programming technologies they have skills on. The first paper's author also assessed the experience of these developers as Web developers, by calculating the number of years on jobs directly related to Web programming. Figure \ref{fig:violin_experience} shows a violin plot with the results. As we can see, 50\% of the developers predicted as experts\ have more than four years of experience on Web-related jobs. 

We reproduced this analysis with \smongodb\ and \ssocketio. For \smongodb, 44 out of 58 developers predicted as experts by the proposed method have pages on Linkedin; for \ssocketio, this happens with 5 out of 10 experts. Furthermore, 28 of such experts (64\%) explicitly mention {\sc MongoDB} on their Linkedin pages; and one developer (20\%) refer to \ssocketio. Therefore, both proportions are lower than the one we reported for \sreact. We claim this happens because \smongodb\ and \ssocketio\ are simple and less complex libraries, when compared with \sreact. For this reason, developers usually do not cite them on Linkedin. For example, one of the experts in \ssocketio\ declare on his GitHub profile that he is one of the library's core developers; but this information is not available on his Linkedin profile.
Due to this reason, we also do not evaluate the years of experience of Linkedin users on \ssocketio\ and \smongodb.

Altogether, this triangulation with Linkedin shows that the proposed clustering-based method was able in most cases to find several GitHub developers with evidences of having experience on the studied libraries. However, before concluding, it is also important to  acknowledge that expertise and experience are distinct concepts; indeed, experience is normally viewed as a necessary condition to achieve expertise~\cite{Ericsson2006, Baltes2018}.

\subsection{Limitations} 
\label{sec:limitations}

Certainly, developers can gain expertise on libraries and frameworks by working on private projects or in projects that are not on GitHub, as highlighted by these developers:\\[-0.35cm]


\noindent{\em None of my projects are publicly on GitHub.} (P037, score 4)\\[-0.35cm]

\noindent{\em My work on GitHub isn’t my strongest. My much larger projects are at work and aren’t open source.} (P503, score 4)\\[-0.35cm]

Thus, the lack of public activity on GitHub is a major obstacle for achieving high recall using approaches like the one proposed in this paper. However, as mentioned before, precision tends to be more important in practical settings than recall. 
If we focus on precision, the proposed clustering approach is effective on identifying experts among GitHub users that frequently contribute to client projects.

To illustrate this discussion, Figure \ref{fig:histograms_react} shows histograms with the percentage of \sreact\ experts in each quintile of the feature distributions (0\%-19\%, 20\%-39\%, etc). We can observe an important concentration of experts in the first and second quintiles, for features like {\em codeChurnClientFiles} (26\%), {\em commitsClientFiles} (37\%), and {\em projects} (57\%). In other words, the histograms  confirm the comments of the survey participants, showing that it is common to have experts with 
sparse activity on GitHub. Indeed, this behavior explains the poor performance of machine learning supervised classifiers in our context, as observed in {\em RQ.1}. By construction, these classifiers predict the  expertise of all developers in the ground truth.
Therefore, the presence of experts at both ends of the distributions showed in Figure~\ref{fig:histograms_react} is a major challenge to their performance. Typically, these classifiers are not able to provide an {\em unknown} answer, as we discussed in Section~\ref{sec:practicalusage}.







\section{Threats to Validity}
\label{sec:threats_validity}




\step{Target Libraries} We mined experts in three popular JavaScript libraries. Thus, it is not possible to fully generalize our findings to experts of other libraries and frameworks.

\step{Candidate Experts} Our list of candidate experts was extracted from an initial list with the top-10K most starred GitHub projects (see Section~\ref{sec:experts}). We acknowledge that our results might be impacted if we expand or reduce this initial list.

\step{Alias Handling} The method used for detecting aliases in the initial list of candidate experts (see Section~\ref{sec:experts}) do not distinguish developers that have multiple GitHub accounts,~i.e., they are considered distinct developers. Therefore, further analysis is required to 
quantify the incidence of such accounts.
  
\step{Ground Truth} Another threat is related to mislabeled classes, due to personal opinions of the surveyed developers, as discussed in Section~\ref{sec:ground-truth}. However, we surveyed 575 developers and some level of mislabeling would not interfere in our results, since the selected algorithms are robust to label noises. Furthermore, to tackle the imbalanced behavior of our ground truth, we used a technique called SMOTE, commonly used on several software engineering problems~\cite{Tan2015,Li2016,Zampetti2017,Zhou2017}. But we acknowledge that there are other techniques , such as over-sampling and cost-sensitive methods~\cite{imbalancedLearning,Chicco2017}.





\section{Related Work}
\label{sec:related_work}
  
 CVExplorer~\cite{Greene2016} is a tool to extract and visualize developers' skills data from GitHub, including skills on programming languages, libraries, and frameworks. The extracted data is presented in the form of a ``tag cloud'' interface, where the  tags denote programming technologies (e.g.,~web development), libraries and frameworks (e.g.,~React) or programming languages (e.g.,~JavaScript). Tags are mined from the project's READMEs and from commit messages. CPDScorer~\cite{Huang2016} is another tool that scores developers' skills, but by correlating developers' activity on Stack Overflow and GitHub. The tool assumes that  
  developers with high quality Stack Overflow answers (measured by number of upvotes) are more likely to be experts in specific programming technologies; the same is assumed for developers who contributed to high quality projects, as measured using source code metrics. 
  Constantinou and Kapitsaki~\cite{Constantinou2016} also propose a repository-mining approach for assessing developer's skills in specific programming languages. Essentially, the aforementioned works differ from the approach described in this paper regarding their methods and goals. CVExplorer considers only commit messages, while we consider the specific files and import statements modified in a commit. CPDScorer works at the level of projects, i.e., the skills acquired by developers on individual commits are not considered. Finally, the approach proposed by Constantinou and Kapitsaki identifies experts in programming languages; by contrast, we target expertise in frameworks and libraries. 


  Hauff and Gousios~\cite{Hauff2015} rely on natural language processing to match job advertisements to GitHub users. First, they extract concept
  vectors from the text of job advertisement and from README's files on GitHub. Then, cosine similarity is used to compare and match these vectors. SCSMiner~\cite{Wan2018} also relies on a vector space model and cosine similarity to calculate the semantic similarity between a project's README and a given query, which can be the name of a programming language or framework or even a more generic skill, such as ``game development''. 
  
  There are also works that rely on machine learning to predict other characteristics and events on software developers life. Wang et. al~\cite{Wang2017} and Mao et al.~\cite{MaoYWJH15} investigate the problem of recommending skilled developers to work on programming tasks posted on the TopCoder crowdsourcing platform. Bao et. al.~\cite{Bao2017} investigate the use of machine learning to predict  developers turn over in two private software companies. 

Lastly, we also identified previous works that approached developers expertise in a more conceptual level. Siegmund et. al.~\cite{Siegmund2014a,Siegmund2014b} asked students a set of questions about their programming experience and then, by means of a controlled experiment, contrasted their answers with the performance of the respondents in program comprehension tasks. They report a strong correlation between the number of tasks successfully concluded and the self-estimates. Baltes and Diehl~\cite{Baltes2018} propose a conceptual framework---obtained from a set of mixed-methods---that maps the main traits around software developers expertise. Their framework reinforces that developers expertise depends on deliberate practice to be enhanced.


  \section{Conclusion}
  \label{sec:conclusion}

Companies often hire based on expertise in libraries and frameworks, as we found in the tags of Stack Overflow jobs. In this paper, we investigated the usage of clustering and machine learning algorithms to identify library experts, using public GitHub data. First, we found that standard machine learning classifiers (e.g., Random Forest and SVM) do not have a good performance in this problem, at least when they are trained with all developers from a sample of GitHub users. The main reason is that not all experts have a strong presence on GitHub. By contrast, we can use clustering techniques to identify experts with high activity on GitHub projects that depend on particular libraries and frameworks. Particularly, we found clusters with 74\% (\sreact), 65\% (\smongodb), and 75\% (\ssocketio) of experts. Supported by these results, we proposed a method to identify library experts based on their similarity (in terms of feature data) to a cluster previously labeled as including a high proportion of experts.

As future work, we recommend to 
(1) investigate other target libraries and frameworks; 
(2) investigate the use of features from other platforms, such as Stack Overflow and TopCoder; and
(3) investigate the accuracy of the proposed method with other developers, including developers of less popular projects.
As a final note, our
data---in a fully anonymized format---and scripts are publicly available at: \url{https://doi.org/10.5281/zenodo.1484498}.

\section*{Acknowledgments}

\noindent We thank the 575 GitHub users who kindly answered our survey. This research is supported by CNPq and FAPEMIG.

\balance

\bibliographystyle{IEEEtran}
\bibliography{IEEEabrv,icsme_2018.bib}

\begin{thebibliography}{10}
\providecommand{\url}[1]{#1}
\csname url@samestyle\endcsname
\providecommand{\newblock}{\relax}
\providecommand{\bibinfo}[2]{#2}
\providecommand{\BIBentrySTDinterwordspacing}{\spaceskip=0pt\relax}
\providecommand{\BIBentryALTinterwordstretchfactor}{4}
\providecommand{\BIBentryALTinterwordspacing}{\spaceskip=\fontdimen2\font plus
\BIBentryALTinterwordstretchfactor\fontdimen3\font minus
  \fontdimen4\font\relax}
\providecommand{\BIBforeignlanguage}[2]{{%
\expandafter\ifx\csname l@#1\endcsname\relax
\typeout{** WARNING: IEEEtran.bst: No hyphenation pattern has been}%
\typeout{** loaded for the language `#1'. Using the pattern for}%
\typeout{** the default language instead.}%
\else
\language=\csname l@#1\endcsname
\fi
#2}}
\providecommand{\BIBdecl}{\relax}
\BIBdecl

\bibitem{Ruiz2014}
I.~J.~M. Ruiz, B.~Adams, M.~Nagappan, S.~Dienst, T.~Berger, and A.~E. Hassan,
  ``A large-scale empirical study on software reuse in mobile apps,''
  \emph{{IEEE} Software}, vol.~31, no.~2, pp. 78--86, 2014.

\bibitem{Sawant2017}
A.~A. Sawant and A.~Bacchelli, ``{fine-GRAPE: fine-grained API usage extractor
  - an approach and dataset to investigate {API} usage},'' \emph{Empirical
  Software Engineering}, vol.~22, no.~3, pp. 1348--1371, 2017.

\bibitem{Treude2011}
C.~Treude, O.~Barzilay, and M.-A. Storey, ``How do programmers ask and answer
  questions on the web? ({NIER} track),'' in \emph{International Conference on
  Software Engineering (ICSE)}, 2011, pp. 804--807.

\bibitem{Siau2010}
K.~Siau, X.~Tan, and H.~Sheng, ``Important characteristics of software
  development team members: an empirical investigation using repertory grid,''
  \emph{Information Systems Journal}, vol.~20, no.~6, pp. 563--580, 2010.

\bibitem{Mockus2002b}
A.~Mockus and J.~D. Herbsleb, ``{Expertise browser: a quantitative approach to
  identifying expertise},'' in \emph{International Conference on Software
  Engineering (ICSE)}, 2002, pp. 503--512.

\bibitem{Fritz2007}
T.~Fritz, G.~C. Murphy, and E.~Hill, ``Does a programmer's activity indicate
  knowledge of code?'' in \emph{Foundations of Software Engineering (FSE)},
  2007, pp. 341--350.

\bibitem{Fritz2010}
T.~Fritz, J.~Ou, G.~C. Murphy, and E.~Murphy-Hill, ``A degree-of-knowledge
  model to capture source code familiarity,'' in \emph{International Conference
  on Software Engineering (ICSE)}, 2010, pp. 385--394.

\bibitem{Fritz2014}
T.~Fritz, G.~C. Murphy, E.~Murphy-Hill, J.~Ou, and E.~Hill,
  ``{Degree-of-knowledge: modeling a developer's knowledge of code},''
  \emph{ACM Transactions on Software Engineering and Methodology}, vol.~23,
  no.~2, pp. 14:1--14:42, 2014.

\bibitem{Schuler2008}
D.~Schuler and T.~Zimmermann, ``Mining usage expertise from version archives,''
  in \emph{International Working Conference on Mining Software Repositories
  (MSR)}, 2008, pp. 121--124.

\bibitem{DaSilva2015}
J.~R. {Da Silva}, E.~Clua, L.~Murta, and A.~Sarma, ``{Niche vs. breadth:
  Calculating expertise over time through a fine-grained analysis},'' in
  \emph{International Conference on Software Analysis, Evolution, and
  Reengineering (SANER)}, 2015, pp. 409--418.

\bibitem{Breiman2001}
L.~Breiman, ``{Random Forests},'' \emph{Machine Learning}, vol.~45, no.~1, pp.
  5--32, 2001.

\bibitem{Weston1998}
J.~Weston and C.~Watkins, ``{Multi-Class Support Vector Machines},'' University
  of London, Tech. Rep., 1998.

\bibitem{DabbishSTH12}
L.~A. Dabbish, H.~C. Stuart, J.~Tsay, and J.~D. Herbsleb, ``{Social Coding in
  {GitHub}: Transparency and Collaboration in an Open Software Repository},''
  in \emph{ACM Conference on Computer-Supported Cooperative Work and Social
  Computing (CSCW)}, 2012, pp. 1277--1286.

\bibitem{ieeesw-2019}
G.~Avelino, L.~Passos, F.~Petrillo, and M.~T. Valente, ``Who can maintain this
  code? assessing the effectiveness of repository-mining techniques for
  identifying software maintainers,'' \emph{IEEE Software}, vol.~1, no.~1, pp.
  1--15, 2019.

\bibitem{icpc2016}
G.~Avelino, L.~Passos, A.~Hora, and M.~T. Valente, ``A novel approach for
  estimating truck factors,'' in \emph{24th International Conference on Program
  Comprehension (ICPC)}, 2016, pp. 1--10.

\bibitem{Singer2013}
L.~Singer, F.~Figueira~Filho, B.~Cleary, C.~Treude, M.-A. Storey, and
  K.~Schneider, ``{Mutual Assessment in the Social Programmer Ecosystem: An
  Empirical Investigation of Developer Profile Aggregators},'' in \emph{ACM
  Conference on Computer-Supported Cooperative Work and Social Computing
  (CSCW)}, 2013, pp. 103--116.

\bibitem{Marlow2013}
J.~Marlow and L.~Dabbish, ``{Activity Traces and Signals in Software Developer
  Recruitment and Hiring},'' in \emph{ACM Conference on Computer-Supported
  Cooperative Work and Social Computing (CSCW)}, 2013, pp. 145--156.

\bibitem{Kruger1999}
J.~Kruger and D.~Dunning, ``Unskilled and unaware of it: how difficulties in
  recognizing one's own incompetence lead to inflated self-assessments.''
  \emph{Journal of personality and social psychology}, vol.~77, no.~6, p. 1121,
  1999.

\bibitem{Siegmund2014a}
J.~Siegmund, C.~K{\"{a}}stner, S.~Apel, C.~Parnin, A.~Bethmann, T.~Leich,
  G.~Saake, and A.~Brechmann, ``{Understanding Understanding Source Code with
  Functional Magnetic Resonance Imaging},'' in \emph{International Conference
  on Software Engineering (ICSE)}, 2014, pp. 378--389.

\bibitem{Yu2003}
L.~Yu and H.~Liu, ``{Feature Selection for High-dimensional Data: A Fast
  Correlation-based Filter Solution},'' in \emph{International Conference on
  Machine Learning (ICML)}, 2003, pp. 856--863.

\bibitem{chen2005}
Z.~Chen, T.~Menzies, D.~Port, and B.~Boehm, ``{Finding the Right Data for
  Software Cost Modeling},'' \emph{IEEE Software}, vol.~22, no.~6, pp. 38--46,
  2005.

\bibitem{Bao2017}
L.~Bao, Z.~Xing, X.~Xia, D.~Lo, and S.~Li, ``{Who Will Leave the Company? A
  Large-Scale Industry Study of Developer Turnover by Mining Monthly Work
  Report},'' in \emph{International Conference on Mining Software Repositories
  (MSR)}, 2017, pp. 170--181.

\bibitem{FeatureTransf}
N.~Zumel, J.~Mount, and J.~Porzak, \emph{{Practical Data Science with R}},
  1st~ed.\hskip 1em plus 0.5em minus 0.4em\relax Manning, 2014.

\bibitem{FeatureTransf2013}
M.~Kuhn and K.~Johnson, \emph{{Applied Predictive Modeling}}, 1st~ed.\hskip 1em
  plus 0.5em minus 0.4em\relax Springer, 2013.

\bibitem{Boehm1999}
S.~Chulani, B.~W. Boehm, and B.~Steece, ``{Bayesian Analysis of Empirical
  Software Engineering Cost Models},'' \emph{IEEE Transactions on Software
  Engineering}, vol.~25, no.~4, pp. 573--583, 1999.

\bibitem{Raudys1991}
S.~J. Raudys and A.~K. Jain, ``{Small Sample Size Effects in Statistical
  Pattern Recognition: Recommendations for Practitioners},'' \emph{IEEE
  Transactions on Pattern Analysis and Machine Intelligence}, vol.~13, no.~3,
  pp. 252--264, 1991.

\bibitem{Japkowicz2002}
N.~Japkowicz and S.~Stephen, ``{The Class Imbalance Problem: A Systematic
  Study},'' \emph{Intelligent Data Analysis}, vol.~6, no.~5, pp. 429--449,
  2002.

\bibitem{Chawla2002}
N.~V. Chawla, K.~W. Bowyer, L.~O. Hall, and W.~P. Kegelmeyer, ``{SMOTE:
  Synthetic Minority Over-sampling Technique},'' \emph{Journal of Artificial
  Intelligence Research}, vol.~16, no.~1, pp. 321--357, 2002.

\bibitem{Tan2015}
M.~Tan, L.~Tan, S.~Dara, and C.~Mayeux, ``{Online Defect Prediction for
  Imbalanced Data},'' in \emph{International Conference on Software Engineering
  (ICSE)}, 2015, pp. 99--108.

\bibitem{Li2016}
L.~Li, T.~F. Bissyand{\'e}, D.~Octeau, and J.~Klein, ``{Reflection-aware Static
  Analysis of Android Apps},'' in \emph{IEEE/ACM International Conference on
  Automated Software Engineering (ASE)}, 2016, pp. 756--761.

\bibitem{Zampetti2017}
F.~Zampetti, C.~Noiseux, G.~Antoniol, F.~Khomh, and M.~D. Penta,
  ``{Recommending when Design Technical Debt Should be Self-Admitted},'' in
  \emph{IEEE International Conference on Software Maintenance and Evolution
  (ICSME)}, 2017, pp. 216--226.

\bibitem{Zhou2017}
Y.~Zhou and A.~Sharma, ``{Automated Identification of Security Issues from
  Commit Messages and Bug Reports},'' in \emph{Foundations of Software
  Engineering (ESEC/FSE)}, 2017, pp. 914--919.

\bibitem{gridSearch}
M.~Claesen and B.~D. Moor, ``{Hyperparameter Search in Machine Learning},''
  \emph{Metaheuristics International Conference (MIC)}, pp. 1--5, 2015.

\bibitem{kappa}
J.~R. Landis and G.~G. Koch, ``{The Measurement of Observer Agreement for
  Categorical Data},'' \emph{Biometrics}, vol.~33, no.~1, pp. 159--174, 1977.

\bibitem{Gorla2014}
A.~Gorla, I.~Tavecchia, F.~Gross, and A.~Zeller, ``{Checking app behavior
  against app descriptions},'' in \emph{International Conference on Software
  Engineering (ICSE)}, 2014, pp. 1025--1035.

\bibitem{Arafeen2013}
M.~J. Arafeen and H.~Do, ``{Test case prioritization using requirements-based
  clustering},'' in \emph{International Conference on Software Testing,
  Verification and Validation (ICST)}, 2013, pp. 312--321.

\bibitem{Vassallo2017}
C.~Vassallo, G.~Schermann, F.~Zampetti, D.~Romano, P.~Leitner, A.~Zaidman,
  M.~{Di Penta}, and S.~Panichella, ``{A tale of CI build failures: An open
  source and a financial organization perspective},'' in \emph{International
  Conference on Software Maintenance and Evolution (ICSME)}, 2017, pp.
  183--193.

\bibitem{Ng2000}
A.~Ng, ``{Machine Learning Course (Stanford CS229 Lecture notes)},'' 2000.

\bibitem{Rousseeuw1987}
P.~J. Rousseeuw, ``{Silhouettes: A graphical aid to the interpretation and
  validation of cluster analysis},'' \emph{Journal of Computational and Applied
  Mathematics}, vol.~20, no.~C, pp. 53--65, 1987.

\bibitem{Grissom2005}
R.~J. Grissom and J.~J. Kim, \emph{Effect sizes for research: A broad practical
  approach}.\hskip 1em plus 0.5em minus 0.4em\relax Lawrence Erlbaum, 2005.

\bibitem{Romano2006}
J.~Romano, J.~D. Kromrey, J.~Coraggio, and J.~Skowronek, ``{Appropriate
  statistics for ordinal level data: Should we really be using t-test and
  Cohen’sd for evaluating group differences on the NSSE and other surveys},''
  in \emph{Annual Meeting of the Florida Association of Institutional
  Research}, 2006, pp. 1--33.

\bibitem{Linares2013}
M.~Linares-V{\'a}squez, G.~Bavota, C.~Bernal-C{\'a}rdenas, M.~D. Penta,
  R.~Oliveto, and D.~Poshyvanyk, ``{API} change and fault proneness: a threat
  to the success of {Android} apps,'' in \emph{Foundations of Software
  Engineering (FSE)}, 2013, pp. 477--487.

\bibitem{Tian2015}
Y.~Tian, M.~Nagappan, D.~Lo, and A.~E. Hassan, ``What are the characteristics
  of high-rated apps? a case study on free android applications,'' in
  \emph{International Conference on Software Maintenance and Evolution
  (ICSME)}, 2015, pp. 301--310.

\bibitem{Ericsson2006}
K.~A. Ericsson, \emph{The Cambridge Handbook of Expertise and Expert
  Performance}, 2006, ch.~38, pp. 683--704.

\bibitem{Baltes2018}
S.~Baltes and S.~Diehl, ``{Towards a Theory of Software Development
  Expertise},'' in \emph{Foundations of Software Engineering (FSE)}, 2018, pp.
  1--14.

\bibitem{imbalancedLearning}
H.~He and E.~A. Garcia, ``{Learning from Imbalanced Data},'' \emph{IEEE
  Transactions on Knowledge and Data Engineering}, vol.~21, no.~9, pp.
  1263--1284, 2009.

\bibitem{Chicco2017}
D.~Chicco, ``{Ten Quick Tips for Machine Learning in Computational Biology},''
  \emph{BioData Mining}, vol.~10, no.~1, pp. 1--35, 2017.

\bibitem{Greene2016}
G.~J. Greene and B.~Fischer, ``{CVExplorer: Identifying Candidate Developers by
  Mining and Exploring Their Open Source Contributions},'' in \emph{IEEE/ACM
  International Conference on Automated Software Engineering (ASE)}, 2016, pp.
  804--809.

\bibitem{Huang2016}
W.~Huang, W.~Mo, B.~Shen, Y.~Yang, and N.~Li, ``{CPDScorer: Modeling and
  Evaluating Developer Programming Ability across Software Communities},'' in
  \emph{Software Engineering and Knowledge Engineering Conference (SEKE)},
  2016, pp. 01--06.

\bibitem{Constantinou2016}
E.~Constantinou and G.~M. Kapitsaki, ``{Identifying Developers' Expertise in
  Social Coding Platforms},'' in \emph{Euromicro Conference on Software
  Engineering and Advanced Applications (SEAA)}, 2016, pp. 63--67.

\bibitem{Hauff2015}
C.~Hauff and G.~Gousios, ``{Matching GitHub Developer Profiles to Job
  Advertisements},'' \emph{Working Conference on Mining Software Repositories
  (MSR)}, pp. 362--366, 2015.

\bibitem{Wan2018}
Y.~Wan, L.~Chen, G.~Xu, Z.~Zhao, J.~Tang, and J.~Wu, ``{SCSMiner: Mining Social
  Coding Sites for Software Developer Recommendation with Relevance
  Propagation},'' \emph{World Wide Web}, pp. 1--21, 2018.

\bibitem{Wang2017}
Z.~Wang, H.~Sun, Y.~Fu, and L.~Ye, ``{Recommending Crowdsourced Software
  Developers in Consideration of Skill Improvement},'' in \emph{IEEE/ACM
  International Conference on Automated Software Engineering (ASE)}, 2017, pp.
  717--722.

\bibitem{MaoYWJH15}
K.~Mao, Y.~Yang, Q.~Wang, Y.~Jia, and M.~Harman, ``{Developer Recommendation
  for Crowdsourced Software Development Tasks},'' in \emph{IEEE Symposium on
  Service-Oriented System Engineering (SOSE)}, 2015, pp. 347--356.

\bibitem{Siegmund2014b}
J.~Siegmund, C.~K{\"{a}}stner, J.~Liebig, S.~Apel, and S.~Hanenberg,
  ``{Measuring and modeling programming experience},'' \emph{Empirical Software
  Engineering}, vol.~19, no.~5, pp. 1299--1334, 2014.

\end{thebibliography}

\end{document}